\documentclass[onecolumn, amssymb,nobibnotes,aps,prb,nopacs]{revtex4}

\usepackage{amsmath}
\usepackage{graphicx}
\usepackage{subfig}
\usepackage[justification=RaggedRight,font=footnotesize]{caption}

\usepackage{multirow}

\usepackage{csquotes}
\usepackage{caption}

\usepackage{calc}
\newlength{\depthofsumsign}
\newcommand{\nsum}[1][1.4]{
    \mathop{%
        \raisebox
            {-#1\depthofsumsign+1\depthofsumsign}
            {\scalebox
                {#1}
                {$\displaystyle\sum$}%
            }
    }
}

\begin{document}

\title{Cross-country comparisons of scientific performance \\ by focusing on  post-apartheid South Africa}

\author{S. M. Hosseini Jenab\footnote{Email: Mehdi.Jenab@nwu.ac.za}}
\affiliation{Centre for Space Research, North-West University, Potchefstroom Campus,
Private Bag X6001, Potchefstroom 2520, South Africa}

\author{}
\affiliation{North-West University, Potchefstroom Campus,
Private Bag X6001, Potchefstroom 2520, South Africa}

\date{\today}

\begin{abstract}
This paper examines the scientific performance of South Africa since 1994 (post-apartheid) 
until 2014 in comparisons with the rest of the world, utilizing relative indicator. 
It provides a view over current standing of South Africa in the scientific world 
as well as its temporal evolution after the apartheid. 
This study focuses on four major aspects of scientific performance, namely quantity, productivity, impact and quality, 
as the main attributes of scientific perfomance on national level.
These are measured by re-based (relative) 
publication, publication per population or GDP, citations and citations per publication respectively.
The study focuses on scientific outputs (in the form of papers published in peer-reviewed journals) 
and their impact (measured by the citations they have received)
to bring into a light a homogeneous comprehension of South Africa's scientific performance
in all these four aspects.
Indicators are adopted cautiously by considering the measures put forward recently 
for scientometrics indicators and their usage in the long-term comparisons studies.
The temporal evolution of these indicators for South Africa are discussed
in the context of three major groups of countries, 
namely African countries, developing countries, and developed (including BRICS) countries.
It aims to examine the process of transition of South Africa from a developing world economy system
into a knowledge-based and innovation-driven one of the developed world. 
The study reveals that South Africa has shown steady increase
in its scientific performance during the studied period  
when compared to the rest of the world.
However, due to the increasing competition from the other developing countries, 
South Africa's position stands the same during this period, 
while countries such as China, Iran, Turkey and Malaysia have shown 
great jump at least in the quantity of their scientific performance.
Furthermore, the study provides the rank of South Africa globally  in both 1996 (initial year of the studied period)
and 2014 (ending year of the study). 
These ranks, changes in them, and their interpretations are discussed by considering their distributions.
\end{abstract}
\maketitle
\section{Introduction} \label{Sec_Intro}
\subsection{Motivation}
The national level scientometrics has been increasing recently due to a global shift towards innovation-driven economy. 
This tendency calls on measuring scientific performance of counties as a proxy of their success in the new world economy.
As governments, societies and corporations realize the significance  of the new system of economy based on high-tech products, 
human capital and information (OECD 1996)\cite{OECD1996}, the importance of research, education and innovation grows. 
In this era, the competitiveness of a nation relays on the level of their performance in producing knowledge and 
consuming it in the form of manufacturing ultra-modern softwares and hardwares. 
In other words, the assessment of national success in financial terms is closely associated with 
such indicators  as quality and quantity of its scientific performance. 
It is suggested that technological progress can enable a cycle of exponential growth by economists such as Zvi Griliches and Kenneth Arrow.
As  Griliches \cite{griliches1998introduction} emphasized on the relationship between economical growth and scientific performance: 
\begin{displayquote}
 ``Real explanations [of productivity growth] will come from understanding
the sources of scientific and technological advances and from identifying 
the incentives and circumstances that brought them about and that facilitated their implementation and diffusion''
\end{displayquote}
For instance, the near-to-exponential GDP growth of United States during the last 130 years 
has been identified to relate to its heavy investment in science and technology \cite{press2013s}.

This push of financial evolution of human societies is echoed in 
many countries' national road-maps regarding their future science-policies.
For instance, Iran insists on 
\begin{displayquote}
 ``achieving the first place in science and technology within the Islamic world, and obtaining a high-rank scientific
place in the world''
\end{displayquote}
as its first goal in its ``national road-map of science''\cite{Iran2010},
and the European union foresees to become
\begin{displayquote}
``... the most competitive and dynamics knowledge-based economy in the world...''.
\end{displayquote}
South Africa recognized this trend \cite{swilling2014rethinking} in the opening line of its 
``innovation towards a knowledge-based economy 2008-2018'':
\begin{displayquote}
``This \textit{Ten-Year Innovation Plan} proposes to help drive South Africa's transformation 
towards  a  knowledge-based  economy,  in  which  economic  growth  is  lead  by  the  
production  and  dissemination  of  knowledge  for  the  enrichment  of  all  fields  of  
human endeavor''.
\end{displayquote}
This document has set a variety of detailed targets to achieve this goal,
most of them are relative and cross-country in nature;
for example:
\begin{itemize}
 \item ``By 2018, South Africa will have:  2.5 percent global share of research publications''.
 \item ``Human and social dynamics: as a leading voice among developing
countries, South Africa should contribute to a greater global understanding
of shifting social dynamics, and the role of science in stimulating growth and
development.''
\end{itemize}

This shift of policies creates a demand for measurement framework to evaluate countries achievement in pursuing
their goals as well as cross-country evaluation to determine their relative position.
In other words, the judgment over national scientific standing in the world of science plays a crucial role for  
governments, businesses and agencies which should decide on funding priorities and scientific preferences.

\subsection{Brief literature review}
The origins of this line of research can be traced back to 
the English chemist Edward Frankland (1825 - 1899) 
who put together figures showing England to be much behind France and
Germany in chemical research by relying on the relative numbers of papers published in 1866
\footnote{Edward Frankland's comments before the 1872 Devonshire Commission, 
which inquired into the state of scientific education in Great Britain,
are cited in Cardwell, Science in England (cit. n.1), p. 96.}.
The concern over the ``national inferiority'' engulfed both England and France by the end of nineteen century. 
The concept of decline has helped to produce some of the impressive quantitative studies in the history and sociology of science.
However, all these early attempts were short of relative positioning and were just concerned mostly to deliver the
message of decline to these countries\cite{nye1984scientific}.

The first well-received study in this line of research was published in 1994 by May \cite{may1997scientific} 
who looked at a few different countries' share of annual scientific outputs of the world and their citations.
He used measures of relative performance to determine the rank of countries. 
The study covers top 15 countries (based on the total number of publications) in the period of 1981-1994.
He concluded that when measuring research performance on national level, 
country size (which can be measured by gross domestic product -GDP- or population) should be taken into account.
However, May's study lacks the standardization for discipline-specific attributes and hence has been criticized heavily for it.  
Furthermore, the temporal evolution of the indicators is overlooked by averaging over all the years and presenting median values.
King \cite{king2004scientific} address the issue of discipline-specification by distinguishing seven broad fields of research.
His study widens the circle of countries from 15 to 31, and focuses on more recent period, from 1993 to 2002. 
However, it still lacks the understanding of scientific activity as an international endeavors which needs
to take into account all the nations to understand the global changes in the culture of scientific activity.
In other words, although he uses the ``Re-Based'' indicators to remove the effect of disciplines discrepancies, 
there is no re-based or normalization for different years, and hence temporal discrepancies are ignored.
Furthermore, the scarcity of information on the funding spent by different countries on each field restricts 
the comparisons based on normalization over investments. 

Recent studies revealed the dependency of scientific performance of countries on the 
international collaborations\cite{sorensen2016studies}. 
Therefore, the concept of national research performance stays unclear to some extent. 
This poses a conceptual challenge in this line of research,
since publications with international co-authorship can not be 
easily assigned equally for all the countries listed in their affiliations. 
Furthermore, the method of counting, either full or fractional counting, brings its own problems to the table. 
Both approaches are concluded to be flawed since they both ignore that fact that credit should be unevenly
distributed. 
Full counting results in what is know as ``inflationary bias'', while fractional counting 
causes ``equalizing bias'' \cite{hagen2015contributory}.

Adding to the conceptual challenges of the research field, the citation pattern for internationally co-authored papers
differ from the others. 
It has been pointed out in a few studies that international co-authored publications receive greater number of citations
\cite{glanzel2001double,persson2004inflationary}, 
which means that international collaborations not only does blur the meaning of national publication but also
causes bias in the citation patterns of the countries involved in those papers.

\subsection{Measures for using scientometrics indicators}
Due to the explosive increase in the indicators used to measure scientific performances on different levels, 
namely individual researchers, scientific journals, universities or research organizations and countries,
some sort of regulation or at least measures are needed to validate them. 
In the latest attempt, the San Francisco declaration on Research Assessment (DORA)\cite{van2013scientists} 
lays some important recommendations to observer when using indicators to assess scientific performance
\footnote{For more information visit: http://www.ascb.org/dora/}. 
The qualities of an appropriate performance indicators have been discussed in the 
scientometrics community and 
they are still away from being settled \cite{bollen2009principal,zahedi2014well}. 
However, there are some basic intuitions which can be helpful. 
The list below presents those major ones which are observed in this study\cite{gingras2014criteria}:
\begin{enumerate}
 \item Adequacy : an indicator should be strongly related to the property it is supposed to measure.
 \item Sensitivity: an indicator should take into account the intrinsic inertia of the object it is
 measuring
 \item Homogeneity: an indicator should stay homogeneous by addressing all the different dimensions of the 
 measurement independently. 
\end{enumerate}
An indicator is defined as a variable that measures and faithfully represents
a specific property of an object under study \cite{lazarsfeld1958evidence}.
Hence, the properties of the indicator should satisfy our instinct (originating from intuition or prior knowledge)
regarding that concept's properties.
One of the major properties includes the 
inertia of the object, i.e. its resistance to change. 
Appropriate indicator should be able to show the same resistance as the object. 
For instance, consider individual, researchers versus countries.
The former has much less inertia than the later. 
Hence, on the level of individuals the change in their ranking or impact can easily
happen (it might just take one year or two to change the raking of specific researchers). 
However, on national level, due to the massive inertia, the change in relative position of countries
should take decades, and national level indicator should be able to reflect that.

Other than these, two more insightful criteria are proposed by Gervers \cite{gevers2014scientific}:
\begin{enumerate}
 \item Insensitivity: small variation in data should not affect the results and more precisely should not change 
 ranking or classification
 \item Normalization: the data should be normalized to the field, time period and size, or at least the effect 
 of these factors should be noted in any conclusions.
\end{enumerate}
It is vital to emphasize that even a simple comparison of publication of one country in 
two different years which are apart enough is to some extent misleading since
these two numbers are coming from different eras in scientific performance. 
It is something which is widespread in scientometrics studies across different levels. 
One should note that underlying context for scientific activity is changing rapidly and 
performance of 2004 can not easily be compared with 1994 or 2014.
That is why the 
normalization to year plays an essential role to come closer to a correct judgment for cross-year comparisons.
In other words, ``almost inflationary growth of the value of the basic indicators''
(such as publications and citations)
requires relative indicators 
when studying medium or long term trends in bibliometrics \cite{persson2004inflationary,schubert1986relative}.

Numerous factors have changed globally during the studied period,
especially considering the scientific activity. 
A new era of transnational scientific activity has emerged during these years,
and Internet started to impact scientific activity culture heavily \cite{adams2013collaborations}. 
Although a simple method of averaging over years can not replace all the differences for these different years, 
it the first best option to deal with this problem.
In this study, indicators are used cautiously by re-basing them in order to overcome the discipline and temporal discrepancies. 
To do so, all the nations ($> 200$)
\footnote{Note that the number of countries for each fields varies since for small countries there exists years that 
they have not published any paper for some specific fields.}
are included in the averaging to create as little bias as possible in the 
indicators temporal evolutions. 

\subsection{South Africa and its transition}
This study focuses on South Africa as a nation in its post-apartheid era, i.e. from 1996 until 2014. 
After its remarkable peaceful transition from apartheid to democratic regime,
South Africa has experienced a relatively prosperous years. 
It has managed to move from developing world to newly industrialized countries (NIC)\cite{hossain2011panel},
and by joining BRICS group (comprised of Brazil, Russia, India, China and South Africa)
it is guaranteed to be considered as part of transitional countries to the developed world. 
In order to achieve such state, it urgently needs to 
change its economical system into a knowledge-based and high-tech oriented status. 
Although new policies and laws in support of such change have put into place, 
the transition is still seems far-fetched. 
The first necessity for achieving this goal is to boost scientific performance of the country
on a national level to reach the same standards as the developed world. 
Although this transition encompasses different aspects of social and economical life of South Africa, 
this study is mainly concerned about its scientific aspect.
The focus is on the four main aspects of scientific performance, i.e. quantity, productivity, impact and quality. 
These attitudes among the different aspects of the national scientific activity is chosen
to address the main questions of the paper:
What is the standing point (position) of South Africa in the world of science,
especially in the context of Africa and developing world?
The answer to this question will show how far South Africa has come in its transition. 
The study will compare South African scientific performance with the developed world to 
bring lights on the remaining trajectories that South Africa should take to complete the transition. 
Furthermore, the study focuses on the temporal evolution 
in these comparisons in order to show the context within which South Africa is. 
Since the study relies on relative (re-based) indicators, the temporal evolutions of other nations 
is as important as South Africa's itself in this study. 
Specifically talking, 
other successful countries in the developing world have gone through rapid decline/growth during the same time. 
This dynamical context can be reckon when the temporal evolution of South Africa is presented in comparison with other nations. 

\subsection{Four aspects}
Scientific activity on any of the four levels under scientometric study 
(e.g. individuals, journals, universities or groups and nations)
should be considered as a multi-dimensional and complex endeavor. 
Any attempts in measuring it should firstly decide which aspects it is going to focused upon. 
These different aspects can be thought of as layers in which the inner layers
focus on more ambiguous and qualitative aspects. 
Scientific performance is the outermost layer of national scientific activity covering 
the outcomes of the national level scientific system. 
One can discuss the inner layers such as internal structure 
and culture of scientific activity in certain countries.

The internal structure includes concepts such as mobility, collaborations pattern, science policies adopted in the country 
and many other concepts concerning a certain country's scientific system. 
Culture refers to even more ambiguous aspects including educational and research system and its history. 
For example, ``publish or perish'' meme can be found in most of countries nowadays; 
however, its intensity plays a crucial role in the culture of 
national level scientific activity.
Studies on inner layers are limited to a few countries since 
they need to dive deep into those countries' scientific systems and their data bases. 
Hence, the comparisons are challenging to carry out. 
Furthermore, serious limitations are imposed on such studies due to the dearth of data. 
Not to mention the need to be familiar with the context of scientific activity which requires 
a broad vision on the common culture of the countries in the study.

In this study, since our main concern is the relative position of countries, 
the focus is on the outer layer of scientific activity. 
Quantity, productivity, impact and quality are the four major concepts that this study analyzes.
Although the scientific performance is not limited to the four aspects considered here, 
they cover all of its major attributes and
produce a well-defined comprehension of countries position in the world of science.

\section{Methodology} \label{Sec_Method}

In this study, the basic methodological principle used throughout the paper considers 
the papers published in peer-review journals as the unit of assessments. 
The usage of number of publication and citation and other derived indicators 
have been discussed throughly in several publications. 
The inadequacy of these two indicators when comparing different fields and even different years or periods
is a frequent claimed drawbacks \cite{schubert1986relative,schubert1996cross,garfield1979citation}. 
In this study, re-based (relative) indicators are utilized to overcome this problem. 
This method assesses each paper against its own context and standards both in its field and age. 
Since the study focuses on national level scientometrics, statistical reliability can almost be guaranteed. 
It is due to the fact that there are enough papers for each field in each year to make the study statistically plausible. 

When adopting relative indicators in assessment, 
the focus should be on relative standing rather the value of the indicator, 
since the value itself can not reflect any meaningful understanding of the concept. 
Among many other possible approaches, two different methods is used here to make sense of such values. 
Firstly, the value can be assessed against the  expected value for such an indicator. 
However, this causes its own complexity about how such expected value should be calculated. 
Here, in case of three indicators,namely $\overline{PPP}, \overline{PPG}, \overline{CPP}$
the minimum expected value is defined, and comparisons against such values are considered in the 
analysis of the results. 

Secondly, the values can be compared versus each other. 
This basically means looking at the ranking
of the objects under study in regards to that concept. 
Although this is less ambiguous to do,
one should always take into consideration 
if enough and correct number of objects are chosen
for the comparison in order to deliver a fair and comprehensive judgment.
Here, in case of all the indicators this method is adopted. 
The choice of countries to compare South Africa against and the justification behind it  will be discussed
in the next section.
The results of the two approaches is presented in Secs. \ref{Sub_Sec_Quantity}, \ref{Sub_Sec_Productivity}, \ref{Sub_Sec_Impact} and \ref{Sub_Sec_Quality}
for all the four concepts.

The other way to approach this issue is based on sketching the distribution of values. 
Distributions can reveal the discrepancies among the objects and normal or abnormal behaviors. 
It also brings to light the relative position of the object in contrast to the whole ensemble. 
The outcomes of this analysis are presented in Sec. \ref{Sub_Sec_Ranking}.

\subsection{Concepts and Their Indicators}
\subsubsection{Quantity ($\overline{P}$):}
Quantity of scientific performance can be measured in the paper published peer-reviewed journals which 
consist the main output of scientific activity across fields and disciplines\cite{glaser2007social}.
Here, the following formula is used to achieve ``re-based publication'' as quantity indicator.
The head-count method is used to calculate the publications (and citations) number for countries.
First, the re-based publication is calculated for each field in a specific year:
\begin{equation*}
 \overline{pub}(\text{country, year, field}) = \frac{pub(\text{country, year, field})}
					      {\nsum[1.3]\limits_{\text{countries}}{pub(\text{country, year, field})}}.      
\end{equation*}
The value of this relative indicators is basically comparable across years and fields 
since it is standardized against the context of each field in each year. 
Afterwards, they are averaged over different fields to find a single number 
which represents the re-based publication of a country in a given year:
\begin{equation}
 \overline{P}(\text{country, year}) = \frac{\nsum[1.3]\limits_{\text{fields}}\overline{pub}(\text{country, year, field})}
					    {\text{number of fields}}
					    \times 100\%
					    \label{rbp}
\end{equation}

\subsubsection{Productivity ($\overline{PPP}$, $\overline{PPG}$) :}
The concept of productivity refers to the number of outputs
in comparison to input in the form of human capital or financial investments. 
On national level, inputs consists of population and gross domestic product (GDP). 
In this study, both aspects of the concept input are considered.
So, two indicators are studied, namely productivity in financial
term and productivity in terms of human capital. 

Here, the focus is on the productivity of a nation rather than on  the productivity of its scientific community. 
Therefore, the publication is divided by the total population of the country and not just its full-time equivalent (FTE) researchers. 
In other words, the idea is that if a country possesses certain share of the population of the world,
it should be able to produce the same share of scientific outputs of the world.
Hence, publication per population ($PPP$) is used as a proxy to productivity of a country. 
However, according to the consideration outlined in the Introduction, these values of $PPP$ 
can misguide the comparisons. 
Therefore, relative form of $PPP$ is adopted here, based on the following 
definition:
\begin{equation}
 \overline{PPP}(\text{country, year}) = \frac{\overline{P}(\text{country, year})}
					     {\overline{PoP}(\text{country, year})}
					     \times 100\% \label{rbppp}
\end{equation}
in which $\overline{P}$ comes from equation \ref{rbp}.
$\overline{PoP}$ stands for the re-based population, or in other words, the share of population of a specific country 
in a given year of the world population in the same year. 
\begin{equation*}
 \overline{PoP}(\text{country, year}) = \frac{PoP(\text{country, year})}
					     {\nsum[1.3]\limits_{\text{countries}}{PoP(\text{country, year})}}    
\end{equation*}

Following the same logic, one can build the productivity indicator based on financial input. 
Here, the concern is about the productivity of a nation.
So, devision is carried out over total GDP. 
If the publication is divided by the total investment (share of GDP spent on science and research),
then indicator is mostly showing the efficiency of the investment, 
which stands out of the scope of this study.
The formula for this indicator of productivity is as follows:
\begin{equation}
 \overline{PPG}(\text{country, year}) = \frac{\overline{P}(\text{country, year})}
					     {\overline{GDP}(\text{country, year})} 
					     \times 100\%
					     \label{rbppg}
\end{equation}
in which
\begin{equation*}
 \overline{GDP}(\text{country, year}) = \frac{GDP(\text{country, year})}
					{\nsum[1.3]\limits_{\text{countries}}GDP(\text{country, year})}					
\end{equation*}
Note that the values of this indicator is presented in percentage,
and hence
the value $\overline{PPP} = 100\%$ ($\overline{PPG}=100\%$)
presents the case in which the country under study
produce the same share of scientific outputs as its share of population (GDP). 
Such a country has achieved average productivity. 
However, as it will be discussed in Sec.\ref{Sub_Sec_Productivity}, 
most of the developed world show more productivity than average, especially in case of $\overline{PPP}$.

\subsubsection{Impact ($\overline{C}$):}
Citations has frequently been used as a proxy for impact of scientific research \cite{moed2006citation}. 
However, citation, as an indicator (like any other indicators), faces its own restrictions and should be 
used cautiously \cite{macroberts1989problems,nicolaisen2007citation,macroberts2010problems}. 
The major drawback of citation as proxy of scientific impact is the fact that it just measures
one aspect of impact 
(i.e. the impact of publication on other scientists outputs).
Various ways of impact stays out of scope of citation.
For example, readers and the number of those who benefit from reading the paper but do not publish 
a paper in that research line to cite a specific paper.
New instruments have been developed with the help of information technology 
to count all these micro-data; 
however, most of them are limited to 
individual level \cite{priem2012altmetrics,mohammadi2014mendeley,kurtz2010usage}.
Although these new tools have started a renaissance in bibliometrics, 
they can not be utilized for national level due to the complexity of data management.
Hence, this study unwillingly limits the concept of impact to the citations 
(as the available proxy for it) and adopts 
the re-based form of citations.
Firstly, the re-basing is carried out for each field:
\begin{equation*}
 \overline{cite}(\text{country, year, field}) = \frac{cite(\text{country, year, field})}
						      {\nsum[1.3]\limits_{\text{countries}}{cite(\text{country, year, field})}}.
\end{equation*}

Then, the values of re-based citations for different fields are averaged over to produce single number for each country in every year:
\begin{equation}
 \overline{C}(\text{country, year}) = \frac{\nsum[1.3]\limits_{\text{fields}}\overline{cite}(\text{country, year, field})}
					    {\text{number of fields}} \label{rbp}
					    \times 100\%
\end{equation}

\subsubsection{Quality ($\overline{CPP}$):}
Although the definition of quality when it comes to scientific performance
is to some extent vague, and it is conceptually difficult to elucidate, 
a simple and general notion of that is assumed here. 
The impact of a country's output should be at least equal to its share of quantity;
that is to say a country's share of citation should match its share of publication of the world. 
If these two shares are equal to each other,
then the quality of scientific performance
has reached its minimum value (here presented in percentage value $=100\%$). 
Based on this definition, the formula of quality indicator is as follows:
\begin{equation}
 \overline{CPP}(\text{country, year}) = \frac{\overline{C}(\text{country, year})}
					      {\overline{P}(\text{country, year})}
					      \times 100\%
					      \label{rbcpp}
\end{equation}

\subsection{Groups of countries}
As stated in the Introduction, the focus in this study is on the transition
of South Africa from a developing to a developed country, 
and its economy changing into a knowledge-based model. 
Therefore, 
South Africa's scientific performance is compared versus three main groups of countries in the world, e.g.
African, developing and developed (including BRIC countries). 
The overlapping between African and developing countries help to include more countries in the
comparison presented in the figures.

\paragraph*{Africa: }
For each of the indicators, 
a comparison is firstly presented among African countries. 
It is widely believed that South Africa stands as the most prosperous country in Africa, 
and it has moved beyond African nations in terms of economical progress. 
Hence, it is accepted to see South Africa on top of Africa in almost all of the aspects of 
scientific performances. 
However, one should note the recent swift growth among some of African nations 
(some of the fastest growing countries in the world are in Africa nowadays). 
Therefore, it is important to our study to check the relative position of South Africa
within Africa to consider all the movements by other nations in Africa. 
Eight countries which are the most productive in science in Africa are considered.
They include Kenya, Algeria, Tanzania, Egypt, Ethiopia, Nigeria, Tunisia and Morocco. 

\paragraph*{Developing world: }
Secondly and mainly,
South Africa's achievement in scientific performance 
is measured versus developing countries. 
This group consists of eight countries, e.g. Iran, Argentina, Pakistan, Ukraine, Poland, Turkey, Malaysia, Mexico.
The choice of counties relies on the quality of their research 
This collection includes countries 
from all the continents. 
Note that developing countries from Africa are presented in the first group. 
Hence, the first two groups show South Africa in the context of developing world.

\paragraph*{Developed world:}
The next comparison presented for each of indicators 
shows the position of South Africa in the developed world and among BRICS countries, 
including Brazil, Russia, India, China, United States, United Kingdom, Germany, France.

\subsection{Database}
The data, such as number of publications and citations,
needed for the analyses are obtained from the
SCImago Journal and Country Rank portal, which
provides Scopus data arranged according to the country,
subfield of science and year of publication
\footnote{Further information is available at: http://www.scimagojr.com/.}.
The information about population and GDP of countries are collected from the website of 
``The World Bank'' which covers almost all the countries
for the last 40 years \footnote{Further information is available at: http://data.worldbank.org/.}.

\section{Results} \label{Sec_Result}
\subsection{Quantity: Re-Based Publication ($\overline{P}$)} \label{Sub_Sec_Quantity}
\begin{figure}
  \subfloat{\includegraphics[width=1.0\textwidth]{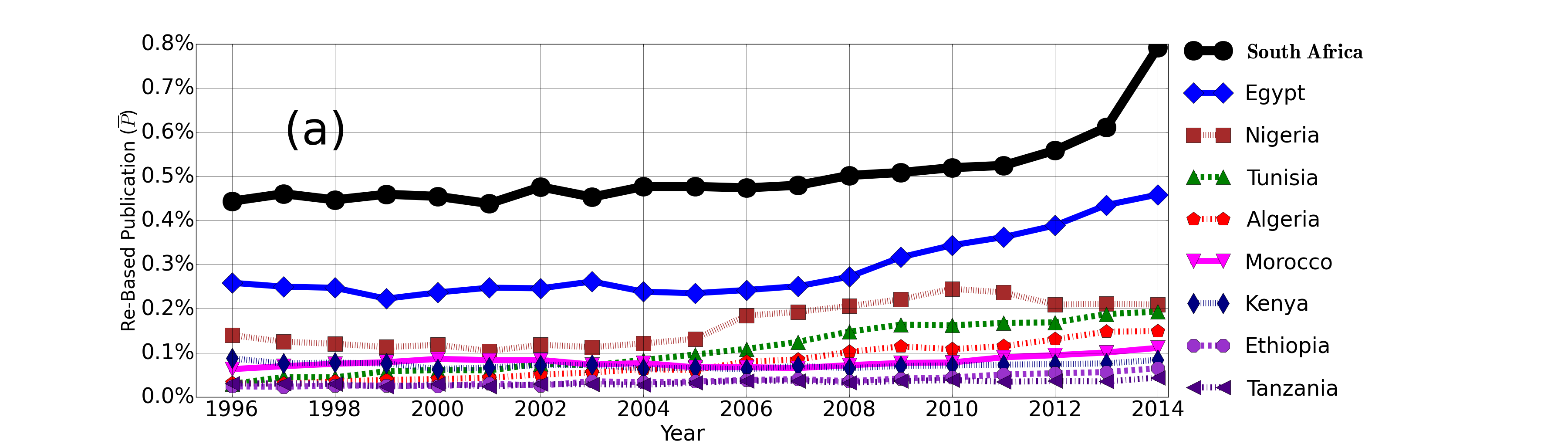}} \\
  \subfloat{\includegraphics[width=1.0\textwidth]{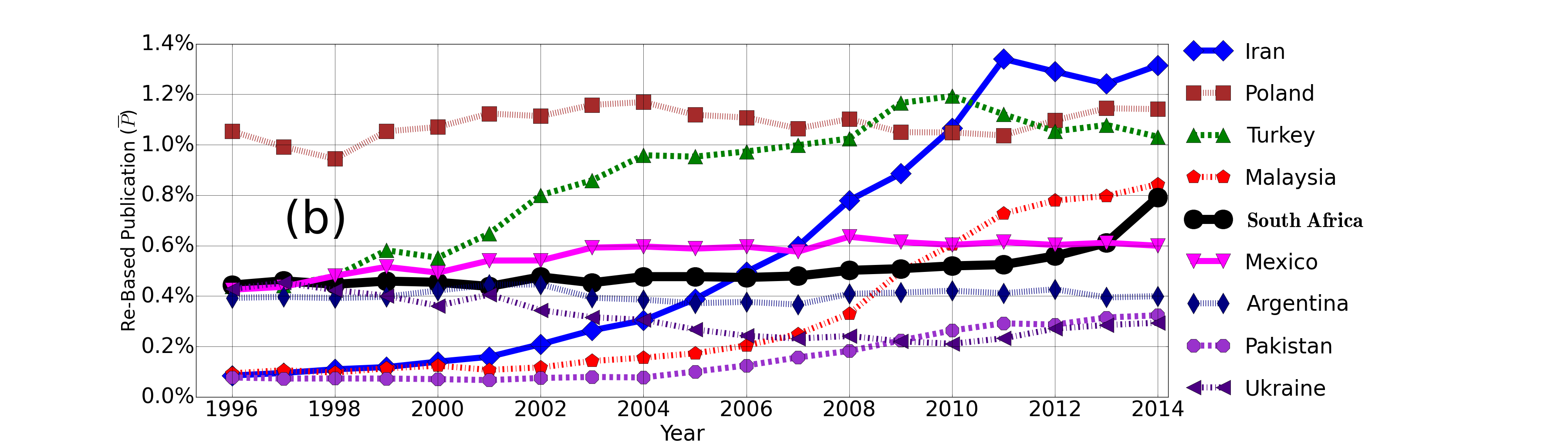}}\\
  \subfloat{\includegraphics[width=1.0\textwidth]{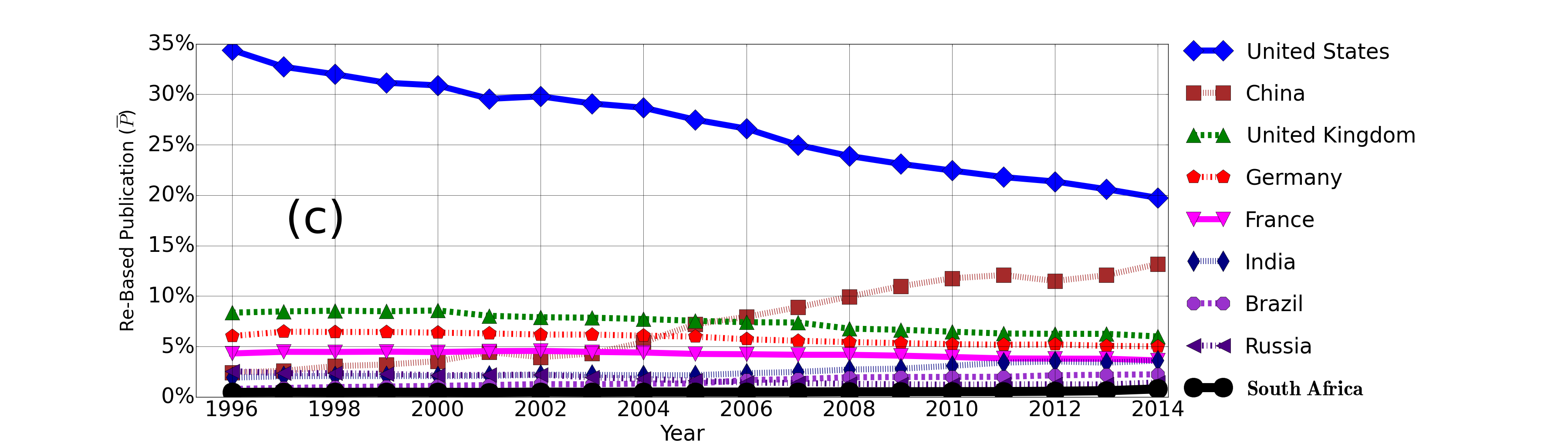}}
  \caption{Comparing Re-Based Publication ($\overline{P}$) of South Africa in the context of (a) Africa,
  (b) developing world, (c) developed world and BRICS group.
  South Africa shows improvement in the last decade and stands the top country in Africa. 
  Among the developing world, it stands as an average country.
  In BRICS countries, South Africa occupies the last position.
  Note the sharp decline of the USA and the rapid growth of countries such as China, Iran, Turkey and Malaysia.}
  \label{RBP}
\end{figure}
Fig. \ref{RBP} presents the comparison in re-based publication of the selected countries with South Africa
in three major groups considered in this study,
namely African countries, developing countries, and developed including BRICS countries. 
It shows that South Africa stands on top of its continent when it comes to the quantity of scientific performance,
followed closely by Egypt. 
Other major African countries by having less than $0.2\%$ of re-based publication of the world
are far below South Africa and Egypt.
In case of Egypt, the relatively close link to the Middle East and Arab world
helps its scientific performance to stay above the rest of Africa. 
South Africa, on the other hand, especially after the apartheid, 
has managed to increase the quantity of its scientific performance by 
adopting new policies focusing on publication as the major form of scientific output\cite{pouris2012science}. 
The incentives policy passed on 2003 which grants scientists considerable amount of payment per publication 
has played a crucial role in this achievement.
This policy insists on scientific publication as an important
source of income for universities in South Africa,
by forcing them to us publication as a ‘carrot and stick’ incentive for their staff. 
Hence, the policy impacts on universities can be summarized as:
\begin{displayquote}
 ``publish or (you and the university will) perish''\cite{kahn2011bibliometric}.
\end{displayquote}
Furthermore, the level of internationalization and 
oversees collaborations has been mentioned 
as the major boost to the scientific activity in South Africa 
\cite{sooryamoorthy2014publication,mouton2000patterns,sooryamoorthy2007does,sooryamoorthy2015transforming}.

However, the considerable growth in re-based publication of South Africa is dwarfed 
by the countries such as Iran, Turkey and Malaysia, from the developing world, 
which are all started from lower than or equal to the position of South African in 1996
and surpassed it by 2014 (see Fig. \ref{RBP}b). 
Most of the developing countries show growth in re-based publication.
Since this indicator is a ``re-based'' one, 
all the improvements of developing world would come at the price of other countries,
mostly developed ones which are experiencing decline in this indicator over studied years. 
This decline is more profound in case of United States 
which has lost half of its share during the studied period, from $35\%$ down to $15\%$.
Among the BRICS countries, South Africa stands with the lowest share in quantity, which reflects the gap
in scientific activity among this group.
China stands way above the rest of the group 
with India, Brazil and Russia standing around the same position
while South Africa is at the bottom. 

This comparison reveals both 
the profound divergence in the quantity of scientific performance around the globe,  
the dynamics and movement that undergoes in scientific performance among the nations. 
The great gap between the USA and the rest of the world has started to close down,
while China managed to secure a strong position
way above the developed countries. 
Among all these fluctuations, 
South Africa was successful to at least 
have growth rate with pace of the average of the developing world.

\subsection{Productivity: Re-Based Publication Per Population ($\overline{PPP}$) 
and Per GDP ($\overline{PPG}$)} \label{Sub_Sec_Productivity}
\begin{figure}
  \subfloat{\includegraphics[width=1.0\textwidth]{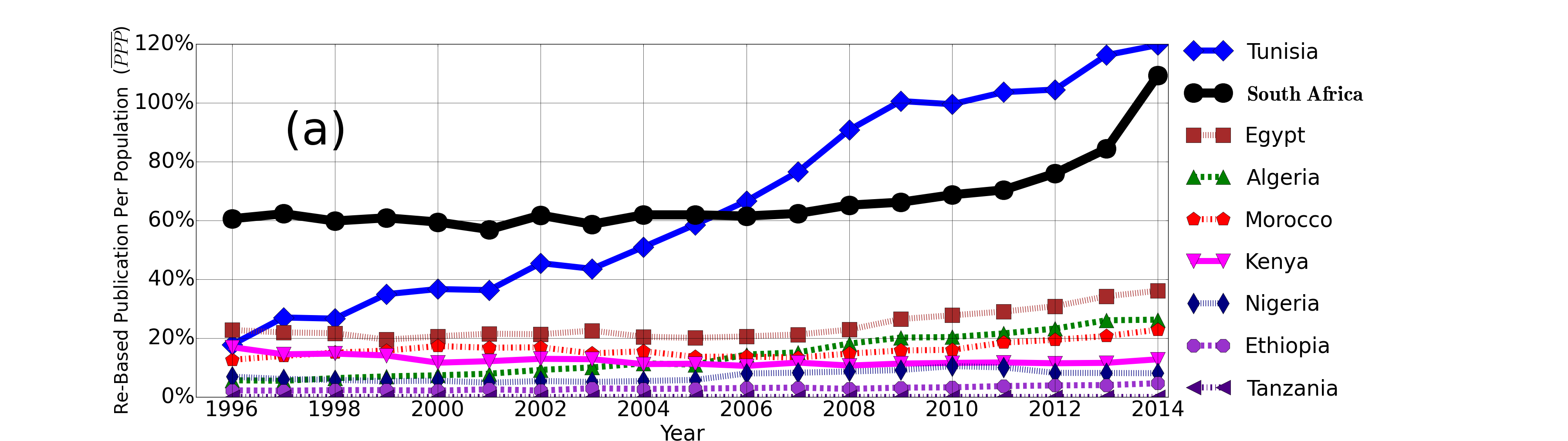}} \\
  \subfloat{\includegraphics[width=1.0\textwidth]{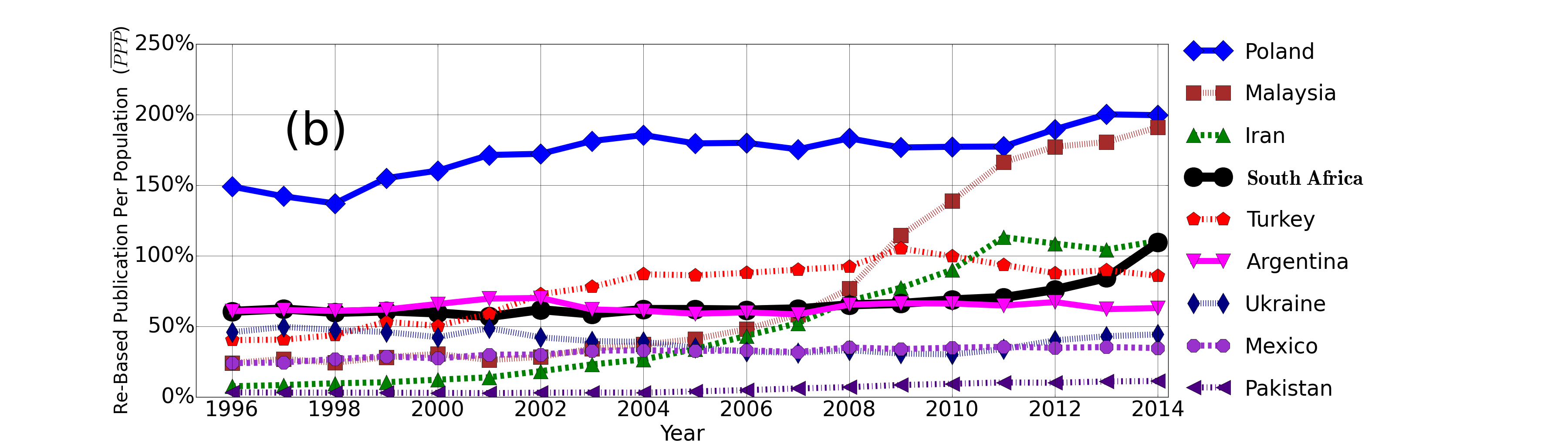}}\\
  \subfloat{\includegraphics[width=1.0\textwidth]{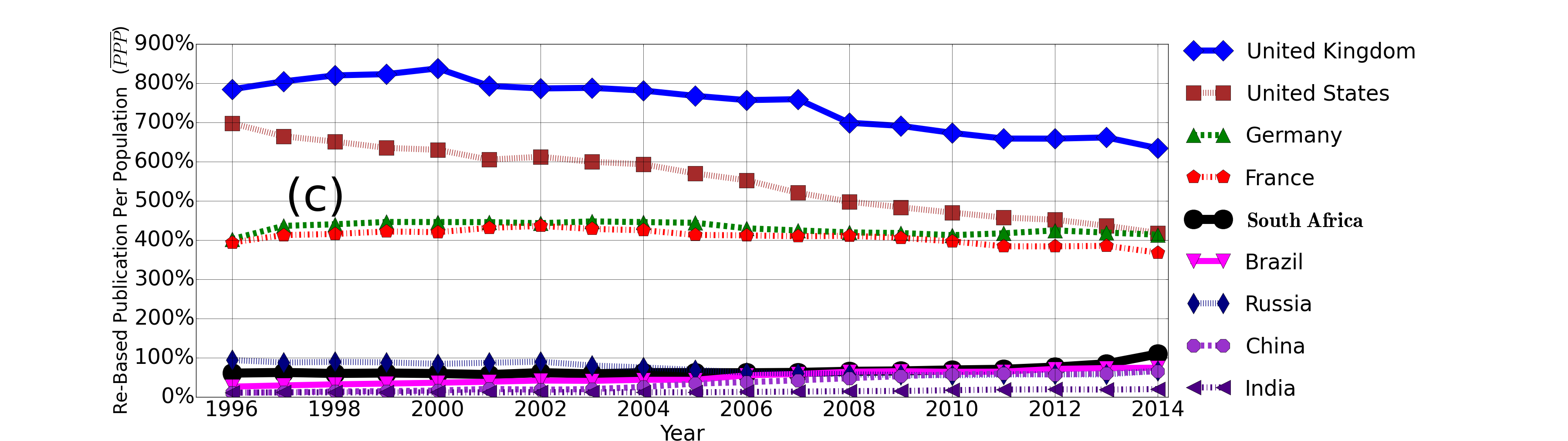}}
  \caption{Comparing Re-Based Publication Per Population ($\overline{PPP}$) for South Africa in the context of 
  (a) Africa, (b) developing world, (c) developed and BRICS countries.
  South Africa shows improvement in the last decade and stands as the second-top country in the continent after Tunisia.
  Most of the developed countries show much higher productivity than the average value ($\overline{PPP} = 100\%$).}
  \label{RBPPP}
\end{figure}

\begin{figure}
\centering
  \subfloat{\includegraphics[width=1.0\textwidth]{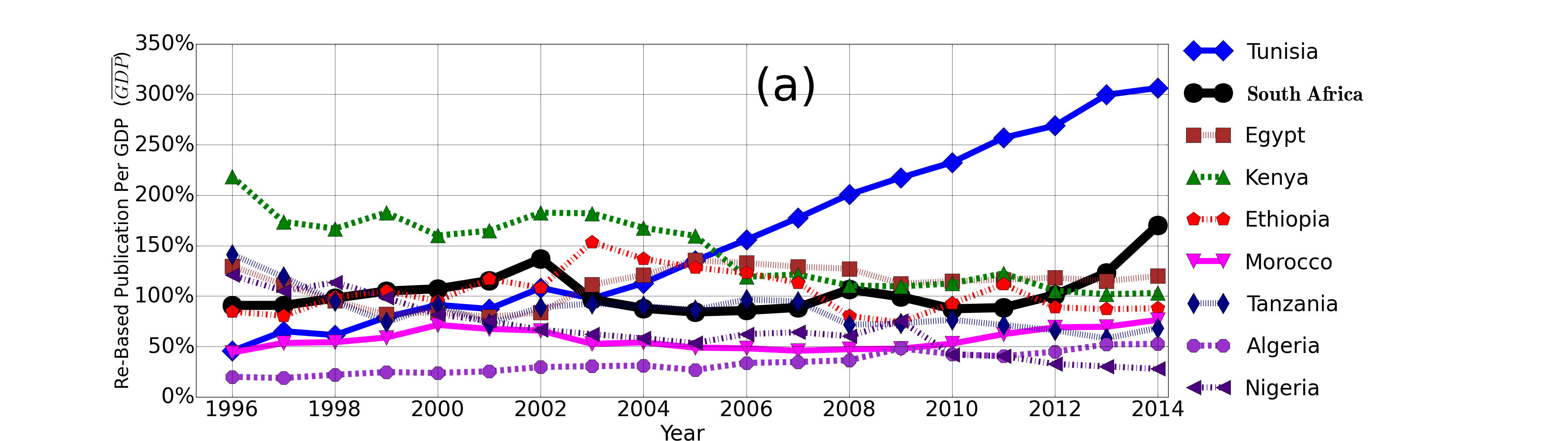}} \\
  \subfloat{\includegraphics[width=1.0\textwidth]{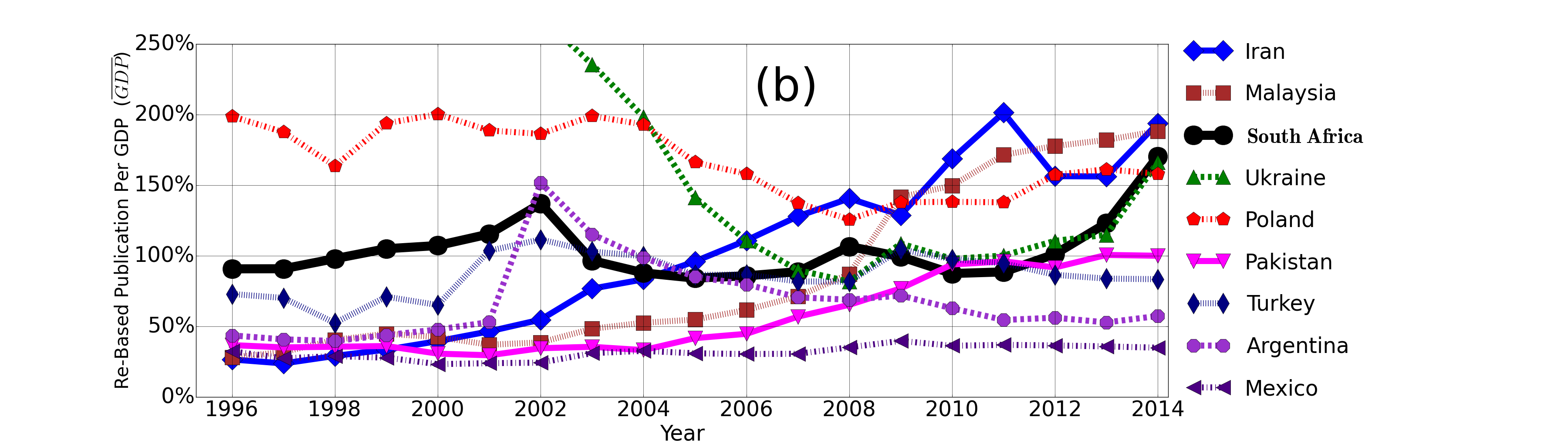}}\\
  \subfloat{\includegraphics[width=1.0\textwidth]{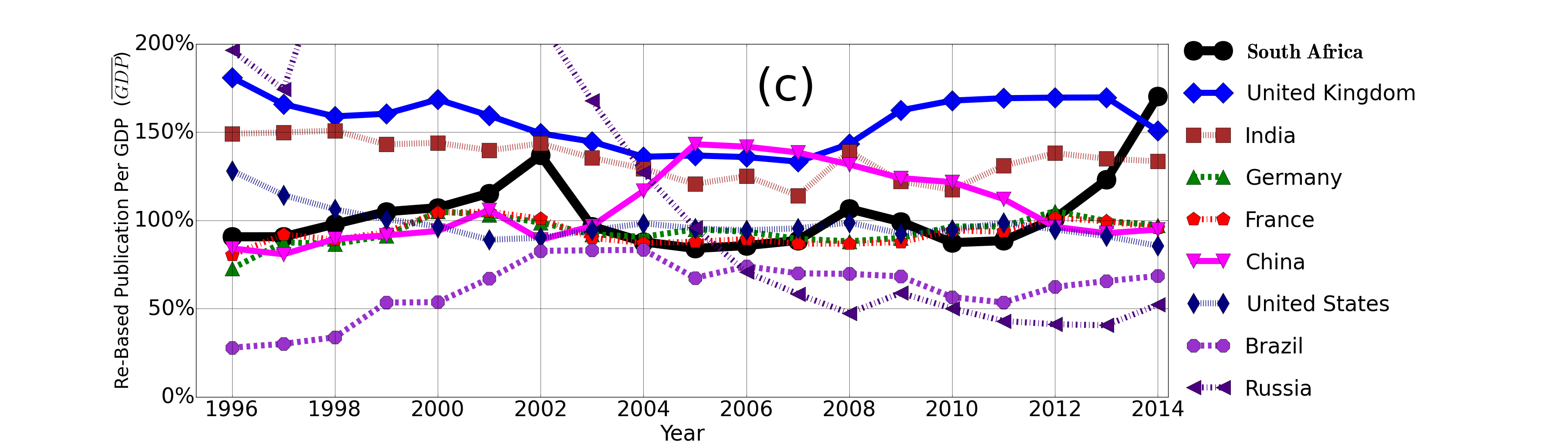}}
  \caption{Comparing Re-Based Citation Per GDP ($\overline{PPG}$) for South Africa in the context of 
  (a) Africa, (b) developing world, (c) developed and BRICS countries.
  South Africa has passed the average value ($\overline{PPG} = 100\%$) just recently.
  Most countries (including the developed ones) stay around the average value.}
  \label{RBPPG}
\end{figure}
Here, both economical and human capital as an input to the scientific system of each countries
have been considered in oder to measure the productivity of the nations. 
The value of $100\%$ stands a threshold, above which countries can be considered as productive.
The idea is that a country should at least have the same share of world publications as its share of population or GDP. 
Figs. \ref{RBPPP} and \ref{RBPPG} present the comparisons for re-based publications per population ($\overline{PPP}$) and per GDP ($\overline{PPG}$)
respectively.

In both these indicators, South Africa shows the productivity near the acceptable minimum of
$\overline{PPP}_{2014} = 80\%$, $\overline{PPG}_{2014} = 120\%$. 
Relatively increasing tendency can be witnessed in both of them when considering the 
temporal evolution after post-apartheid era.
In the context of Africa, Tunisia and South Africa dominate. 
Nevertheless, Tunisia supersedes South Africa in both the indicators by a considerable margin. 
When comparing developing countries, South Africa shows an average productivity;
however,
the temporal evolution of countries such as Iran, Malaysia and Turkey
signals the great changes that have taken place during this era. 
Compared to these countries' big jumps, 
South Africa has not shown that much of development. 
In contrast to the developing world,
South Africa stands as above the rest of BRICS members, which is 
mostly due to the fact that the rest of BRICS countries possess
much larger population or GDP compared to South Africa. 
In conclusion, South Africa's scientific productivity (based on the definition presented here)
shows a minimum acceptable level and has not changed considerably in post-apartheid era, 
which can be worrisome and a source of concern when considering 
the jumps that developing countries has enjoyed in the same era. 

Furthermore, when comparing quantity and productivity,
the polarity appearing in the quantity among the countries reduces dramatically 
in case of productivity, especially GDP-wise comparison. 
This shows that most of the movement and growth visible in the quantity originate
from the disparity in the population or GDP of countries, 
which is a strong evidence of the relationship between financial growth and intensity of scientific activity. 
The growth in the productivity for most of the countries 
are limited compared to some of big jumps in the quantity, 
reflecting the difficulty in boosting productivity of a country. 

\subsection{Impact: Re-Based Citation ($\overline{C}$)} \label{Sub_Sec_Impact}
\begin{figure}
  \subfloat{\includegraphics[width=1.0\textwidth]{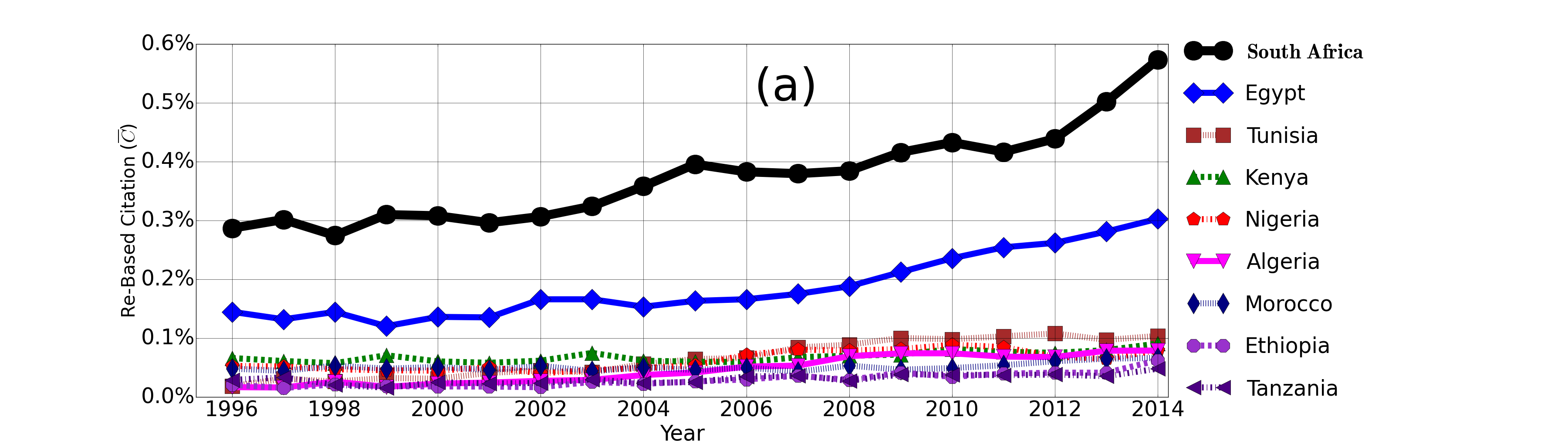}} \\
  \subfloat{\includegraphics[width=1.0\textwidth]{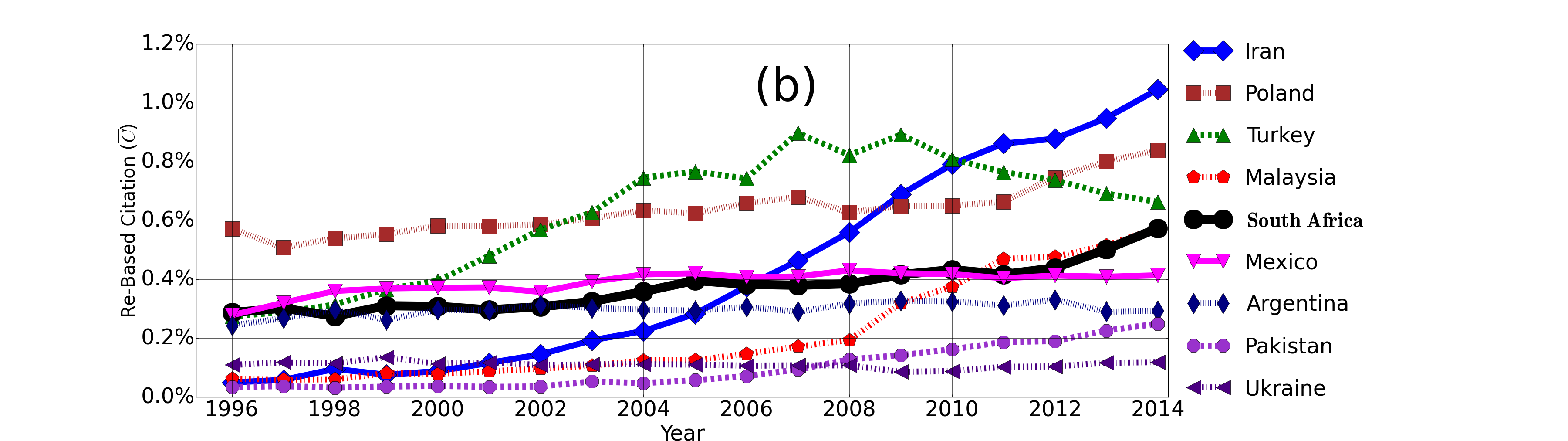}}\\
  \subfloat{\includegraphics[width=1.0\textwidth]{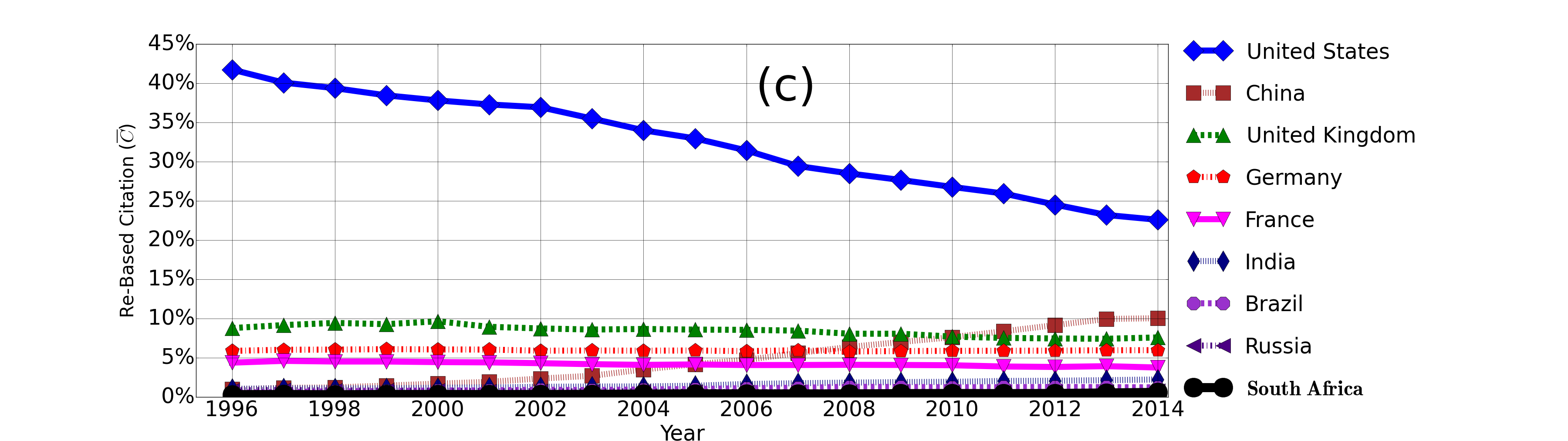}}
  \caption{Comparing Re-Based Citation ($\overline{C}$) for South Africa in the context of 
  (a) Africa, (b) developing world, (c) developed and BRICS countries.
  South Africa enjoys a dominant position in Africa for all the years of the studied period.
  The pattern of decline and growth of countries such as the United States, China, Iran and Malaysia reflect the 
  shift of scientific impact from developed to developing world.}
  \label{RBC}
\end{figure}
Certain dominance of South Africa in the re-based citation is well established among African countries, 
as its achievement during the studied period has always stayed far above the rest of the group (Fig. \ref{RBC}). 
Furthermore, South Africa managed to double its impact during this era. 
However, comparison with developing countries reveals that the overall tendency among these countries 
have grown and in some cases has shown, e.g. Iran, Turkey and Malaysia substantial jumps. 
South African scientific impact on the global level ($\overline{C}_{2014} = 0.6\%$) is still limited ($<1\%$),
as it stands the lowest among BRICS countries and average among developing world. 

United States shows a preeminent impact throughout the studied years; however, decreasing tendency exists, 
even in the last year ($\overline{C}_{2014} = 20\%$)of the study it stands 
twice the value of the next giant China ($\overline{C}_{2014} = 10\%$).

\subsection{Quality: Re-Based Citation Per Publication ($\overline{CPP}$)}\label{Sub_Sec_Quality}
\begin{figure}
  \subfloat{\includegraphics[width=1.0\textwidth]{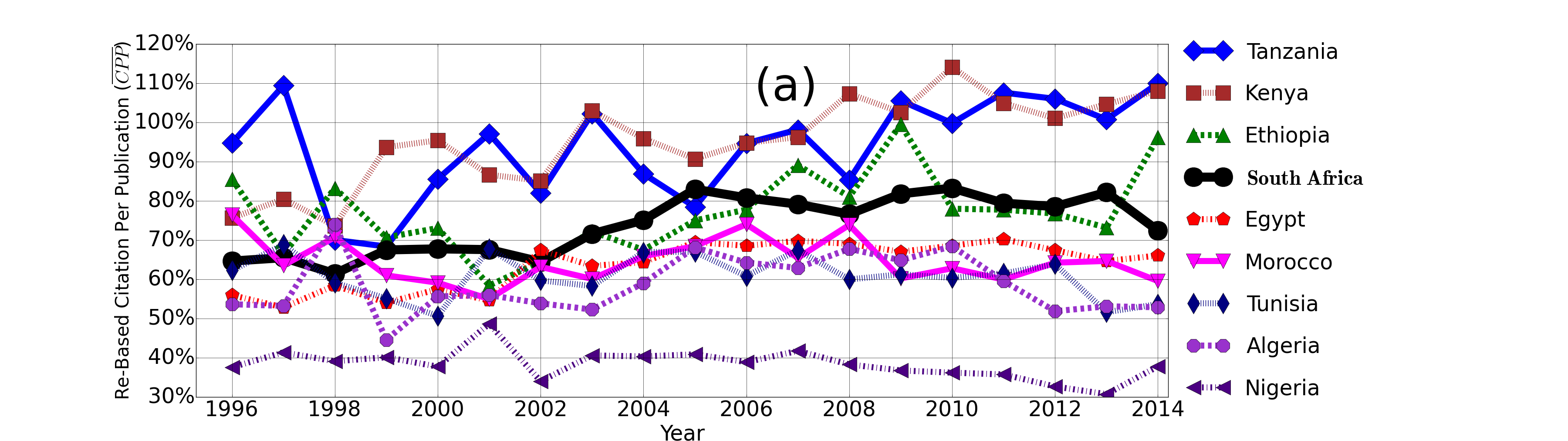}} \\
  \subfloat{\includegraphics[width=1.0\textwidth]{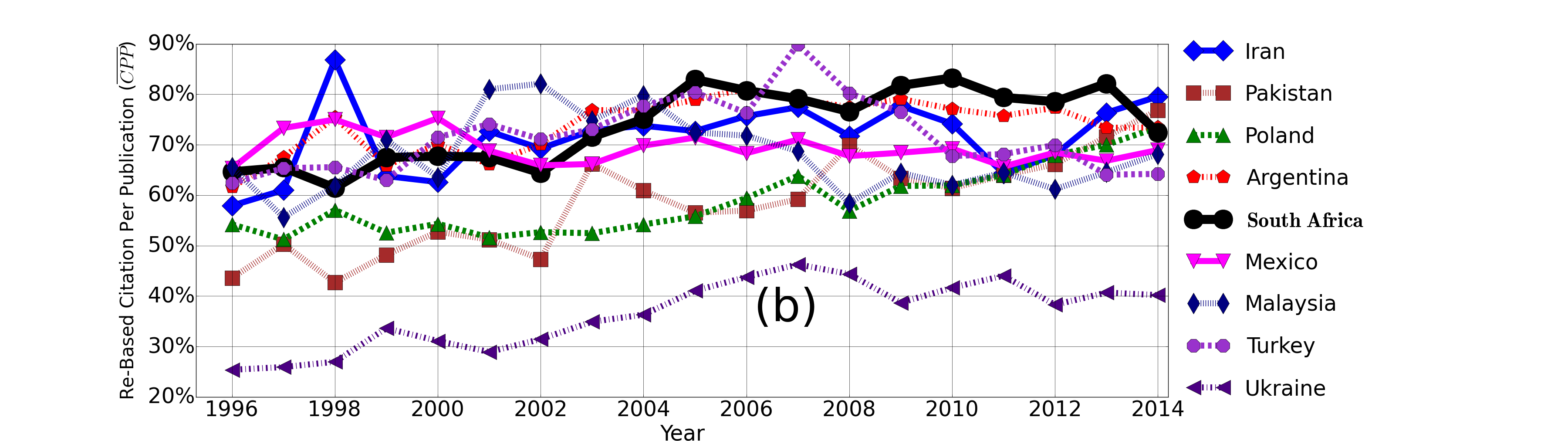}}\\
  \subfloat{\includegraphics[width=1.0\textwidth]{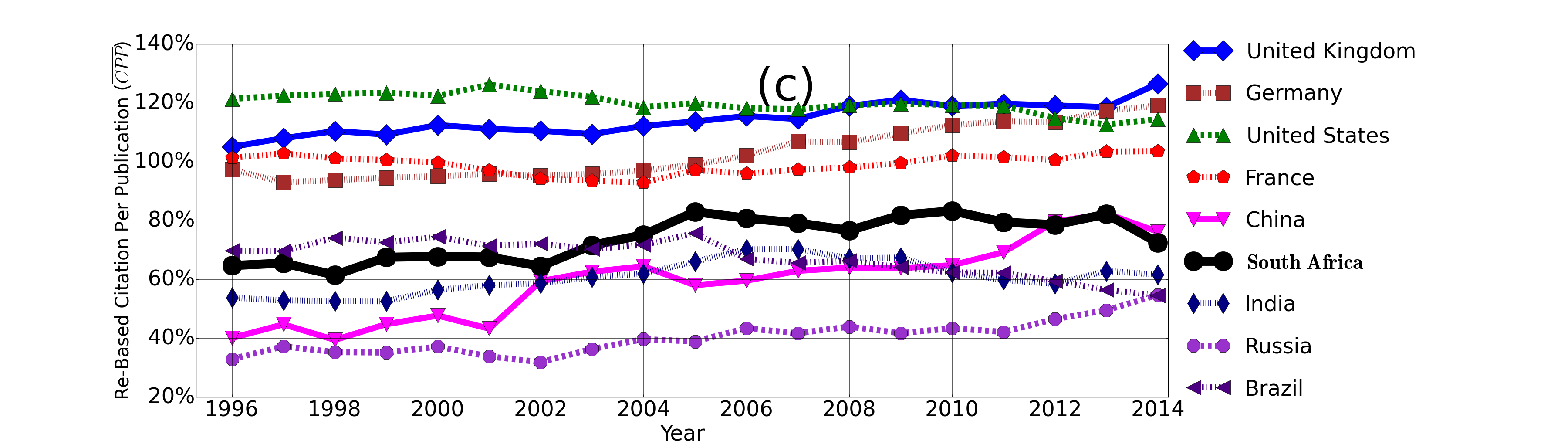}}
  \caption{Comparing Re-Based Citation Per Publication ($\overline{CPP}$) for South Africa in the context of 
  (a) Africa, (b) developing world, (c) developed and BRICS countries.
  In contrast to previous indicators,
  almost all the countries show an steady time line without rapid growth or sharp decline.}
  \label{RBCPP}
\end{figure}
$\overline{CPP} = 100\%$ shows the average quality,
meaning that a country should receive at least the same 
portion of citation of the world as its share of the world's publication. 
South Africa, although growing, remains below the average quality, showing the struggle ahead for 
improving the quality of the research in the country.
The same challenge exists for almost all the developing countries, as most of them show the same 
below-average quality. 

In the context of the developed and BRICS countries,
South Africa stands as the mean value between these two sets of countries. 
Almost all the countries in this group show growth in their quality, especially China, Germany and France. 
In contrast to the quantity of scientific performance of the same group of countries,
it can be witnessed that 
although most of the developed countries are losing ground to the developing world
in the quantity of publications,
their quality stays the 
same or even increases. 

\subsection{Global rankings and distributions} \label{Sub_Sec_Ranking}
\begin{figure}
  \subfloat{\includegraphics[width=0.5\textwidth]{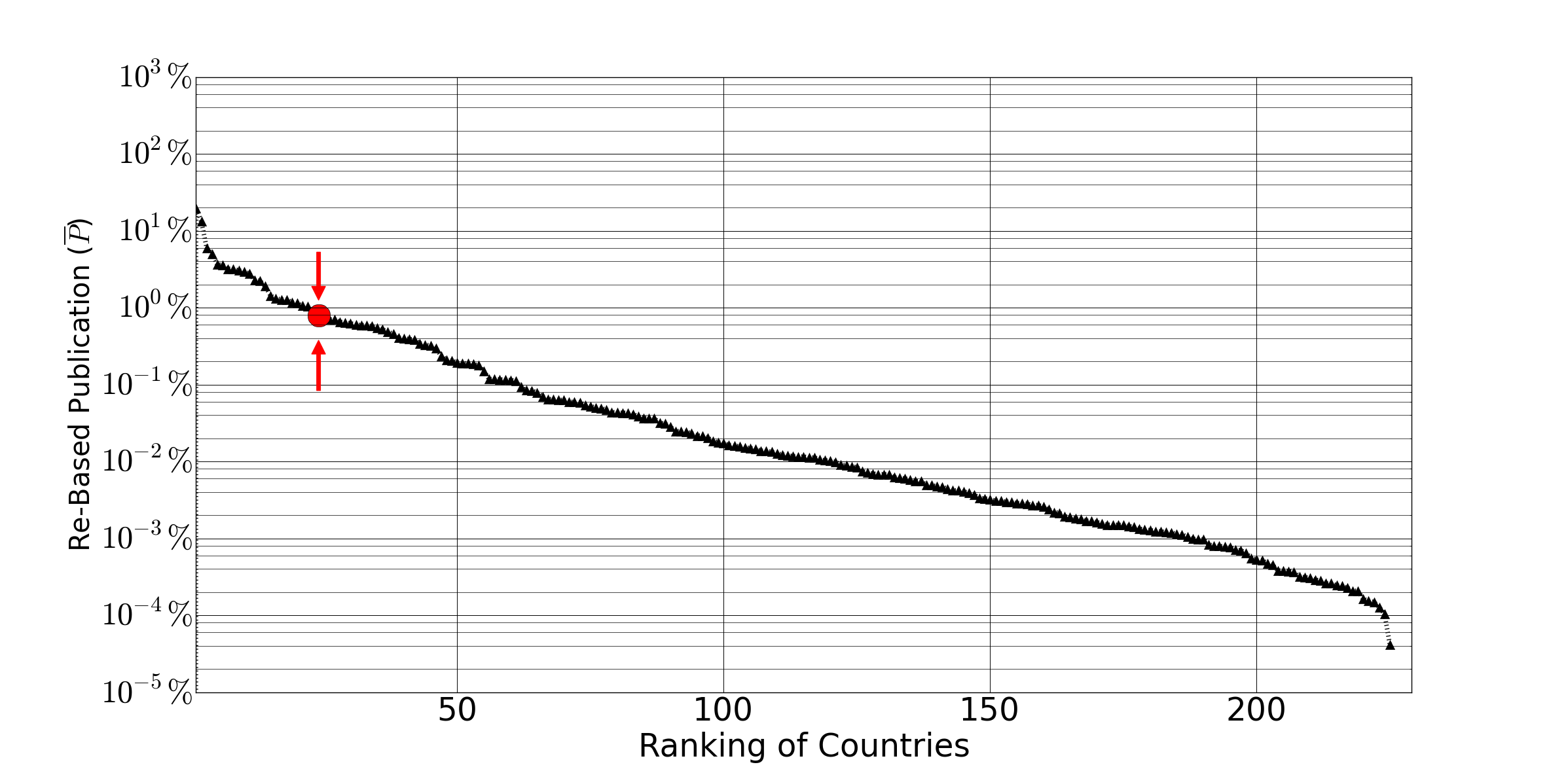}} 
  \subfloat{\includegraphics[width=0.5\textwidth]{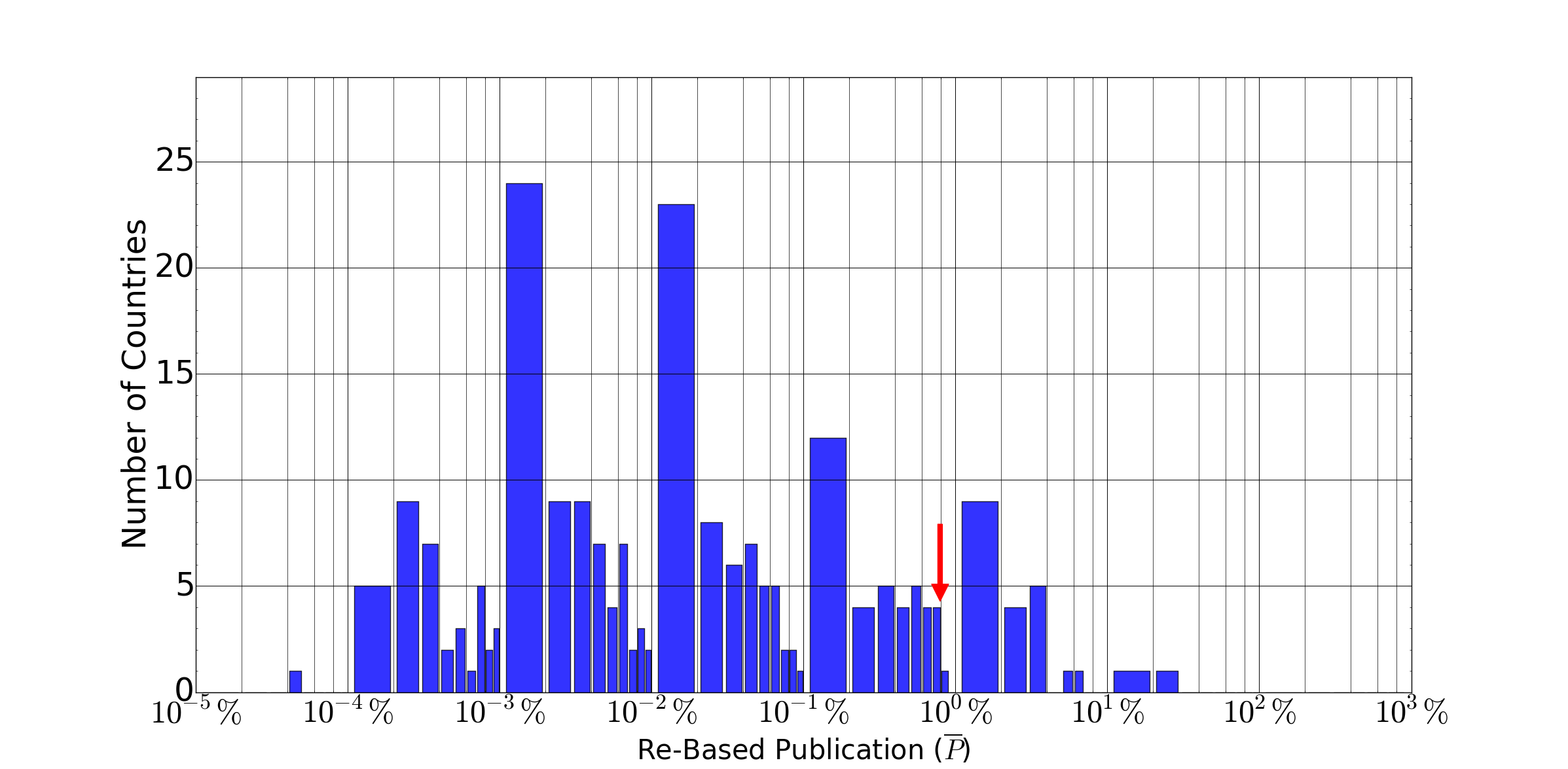}} \\
    \subfloat{\includegraphics[width=0.5\textwidth]{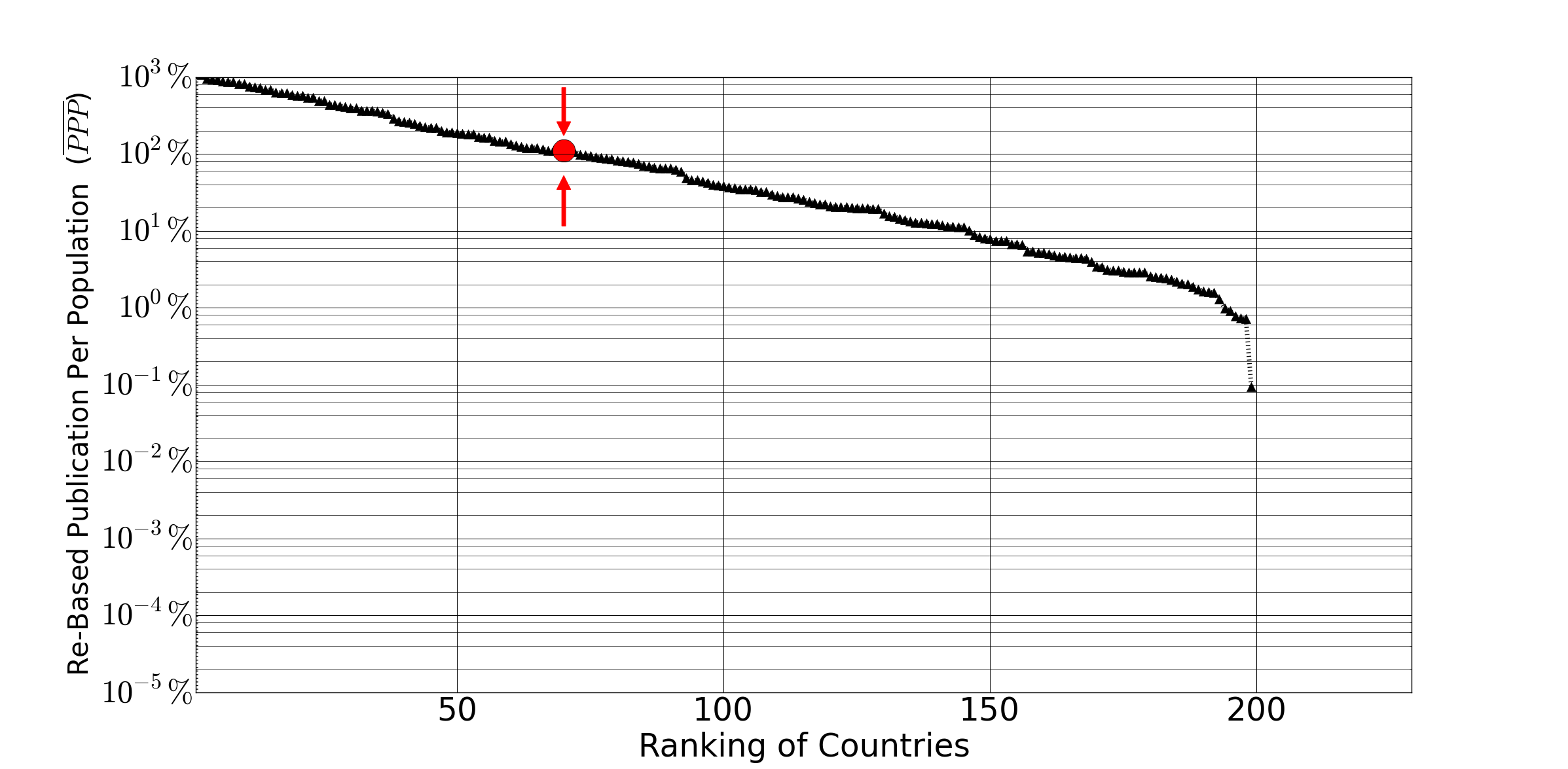}} 
  \subfloat{\includegraphics[width=0.5\textwidth]{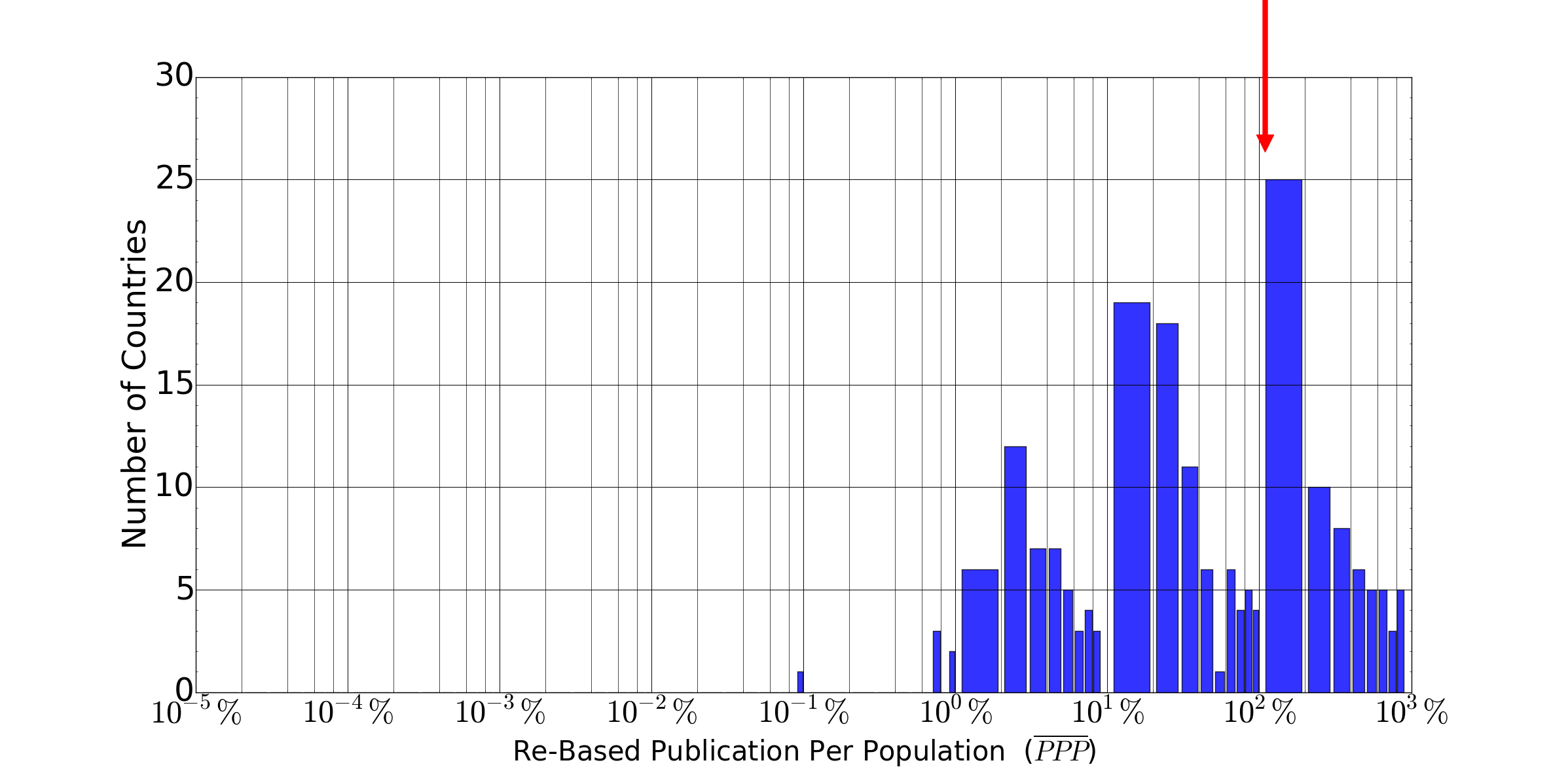}} \\
    \subfloat{\includegraphics[width=0.5\textwidth]{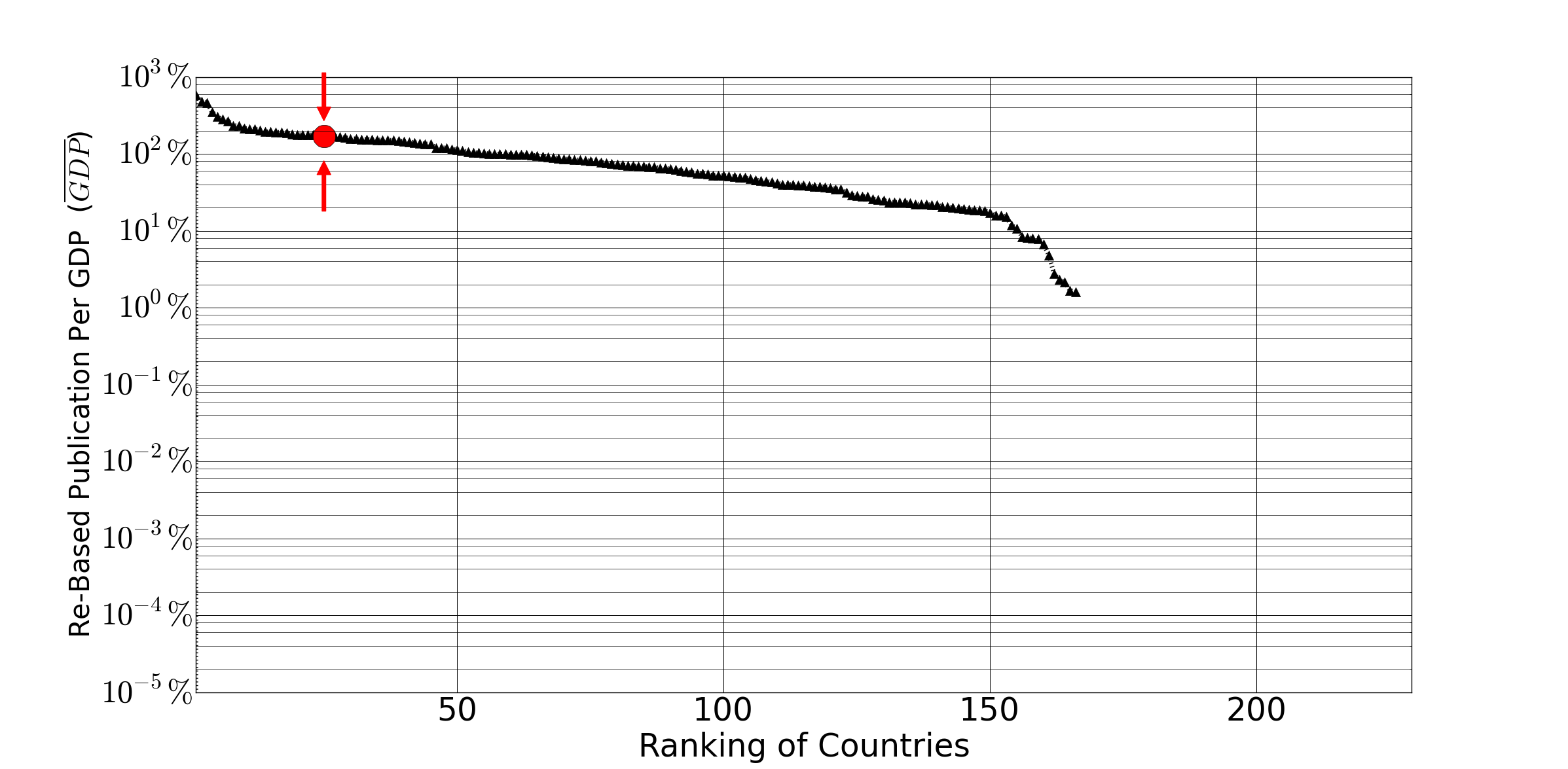}} 
  \subfloat{\includegraphics[width=0.5\textwidth]{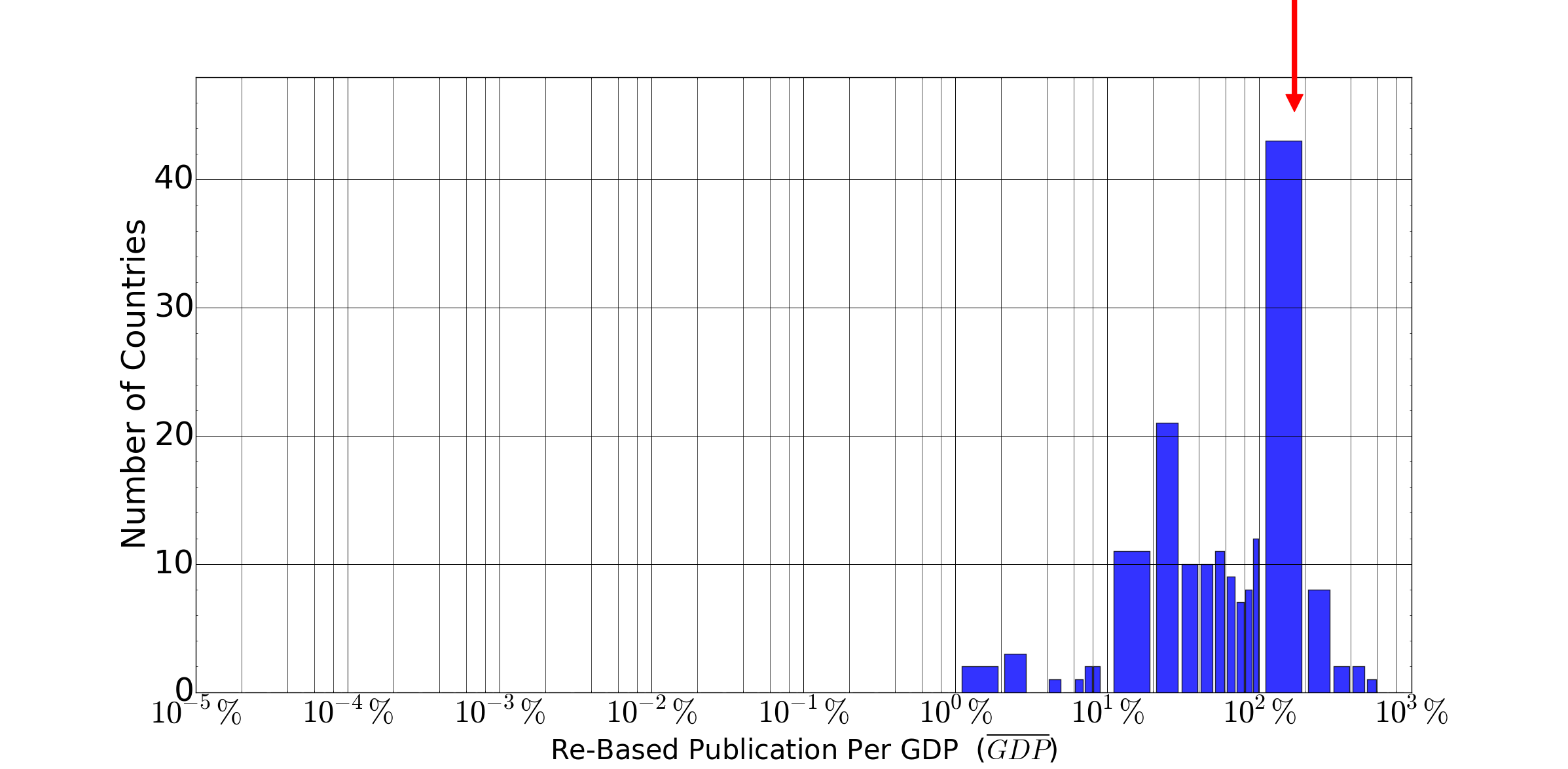}} \\
    \subfloat{\includegraphics[width=0.5\textwidth]{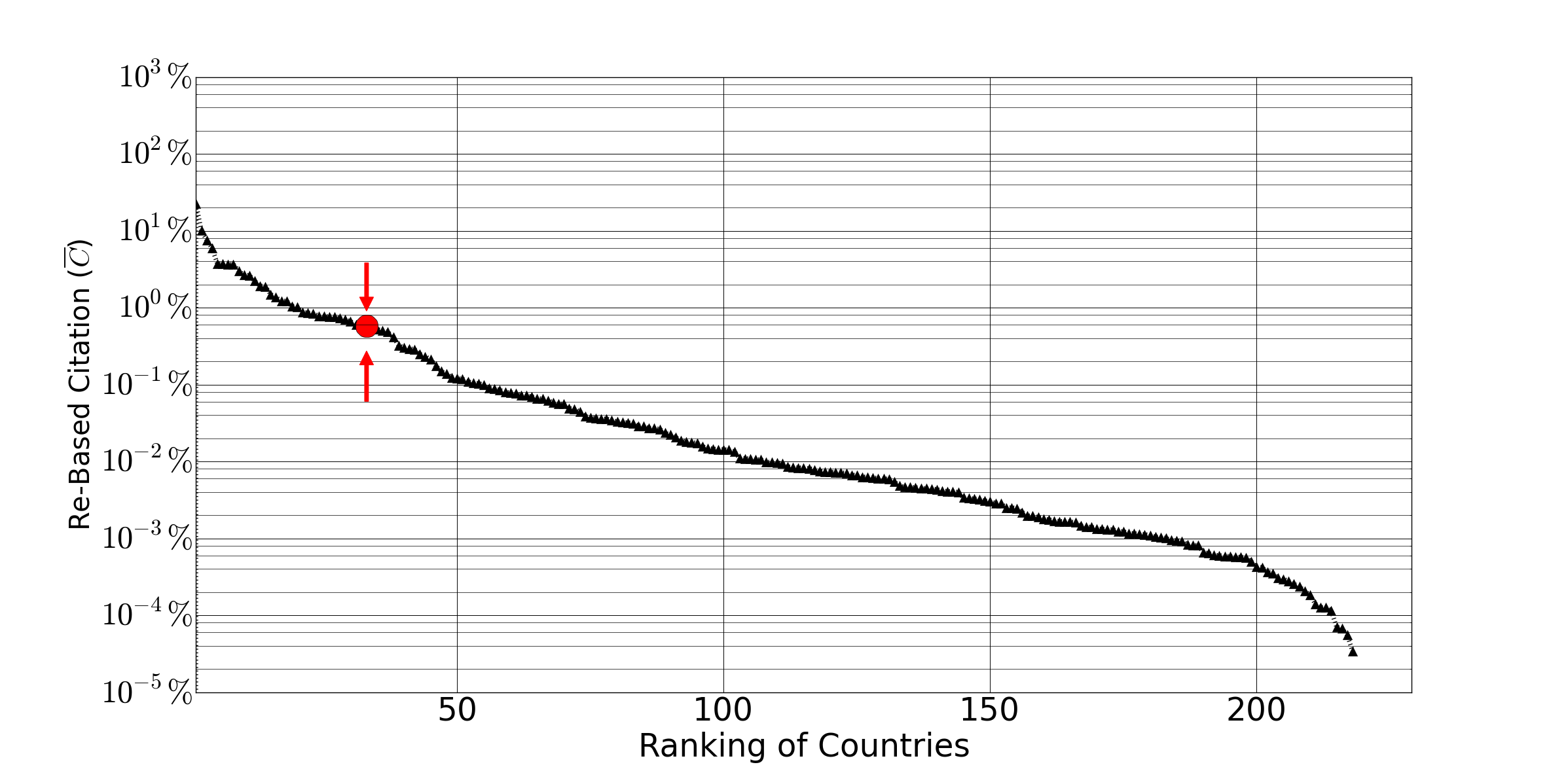}} 
  \subfloat{\includegraphics[width=0.5\textwidth]{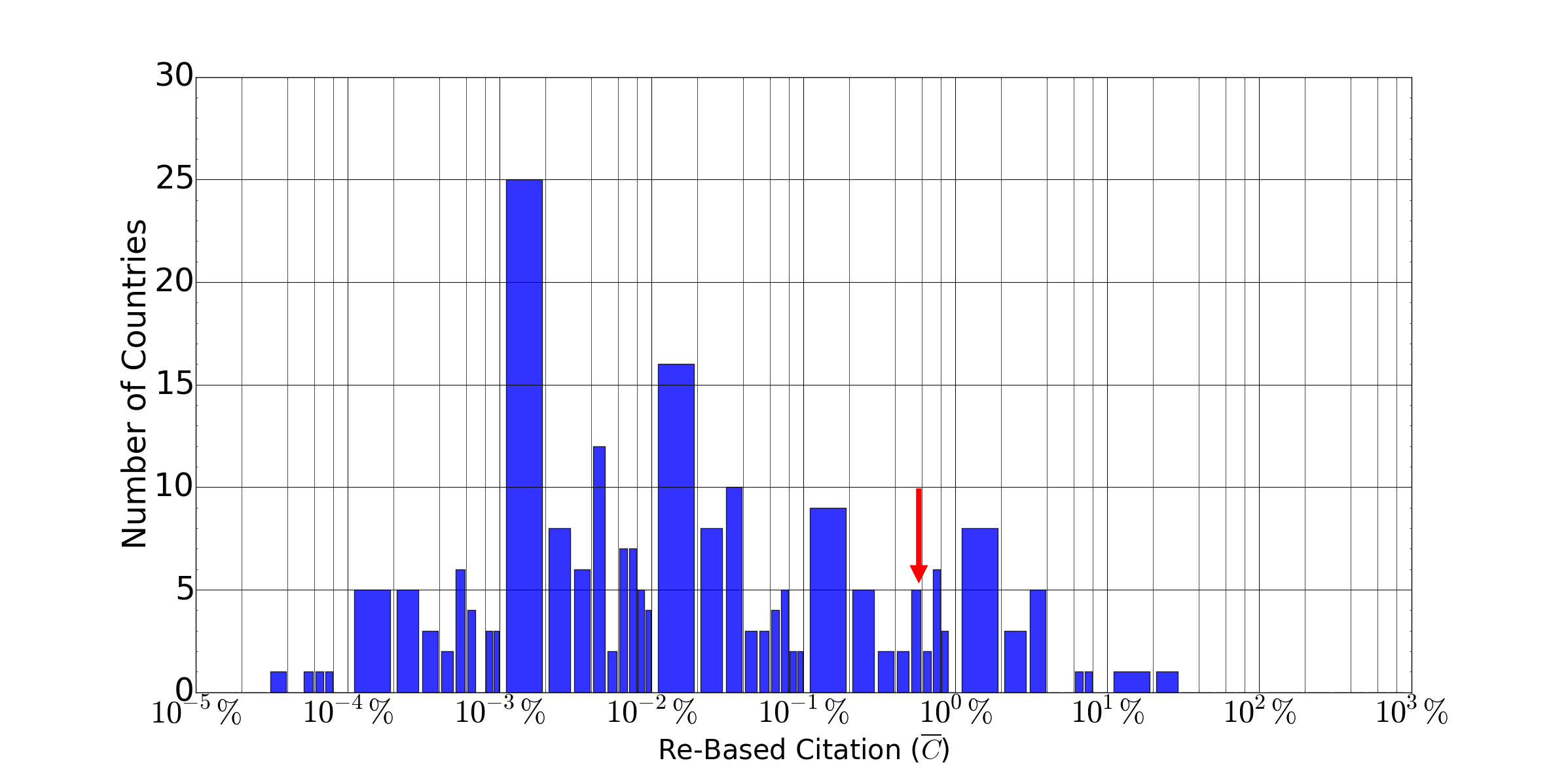}} \\
    \subfloat{\includegraphics[width=0.5\textwidth]{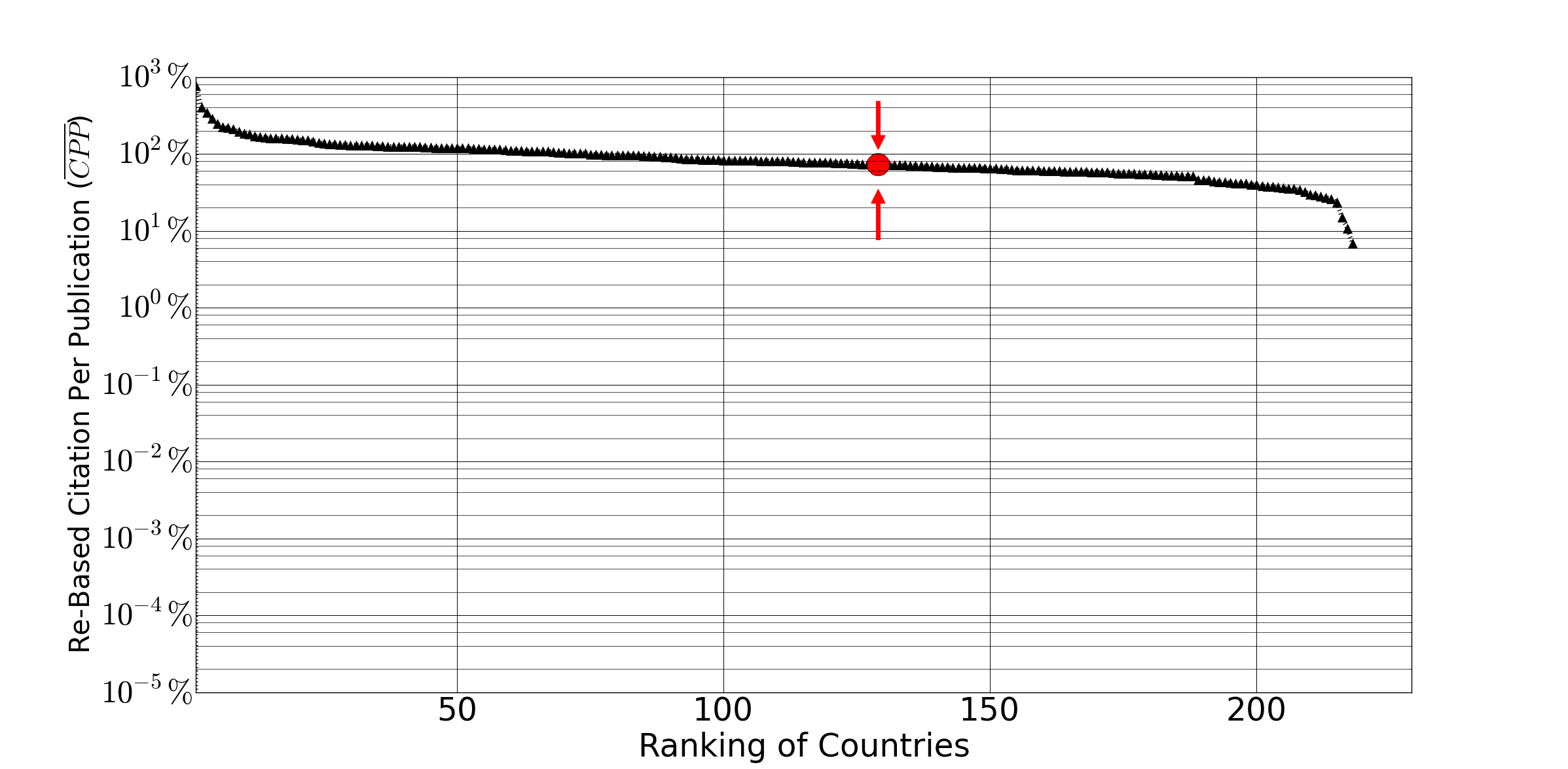}} 
  \subfloat{\includegraphics[width=0.5\textwidth]{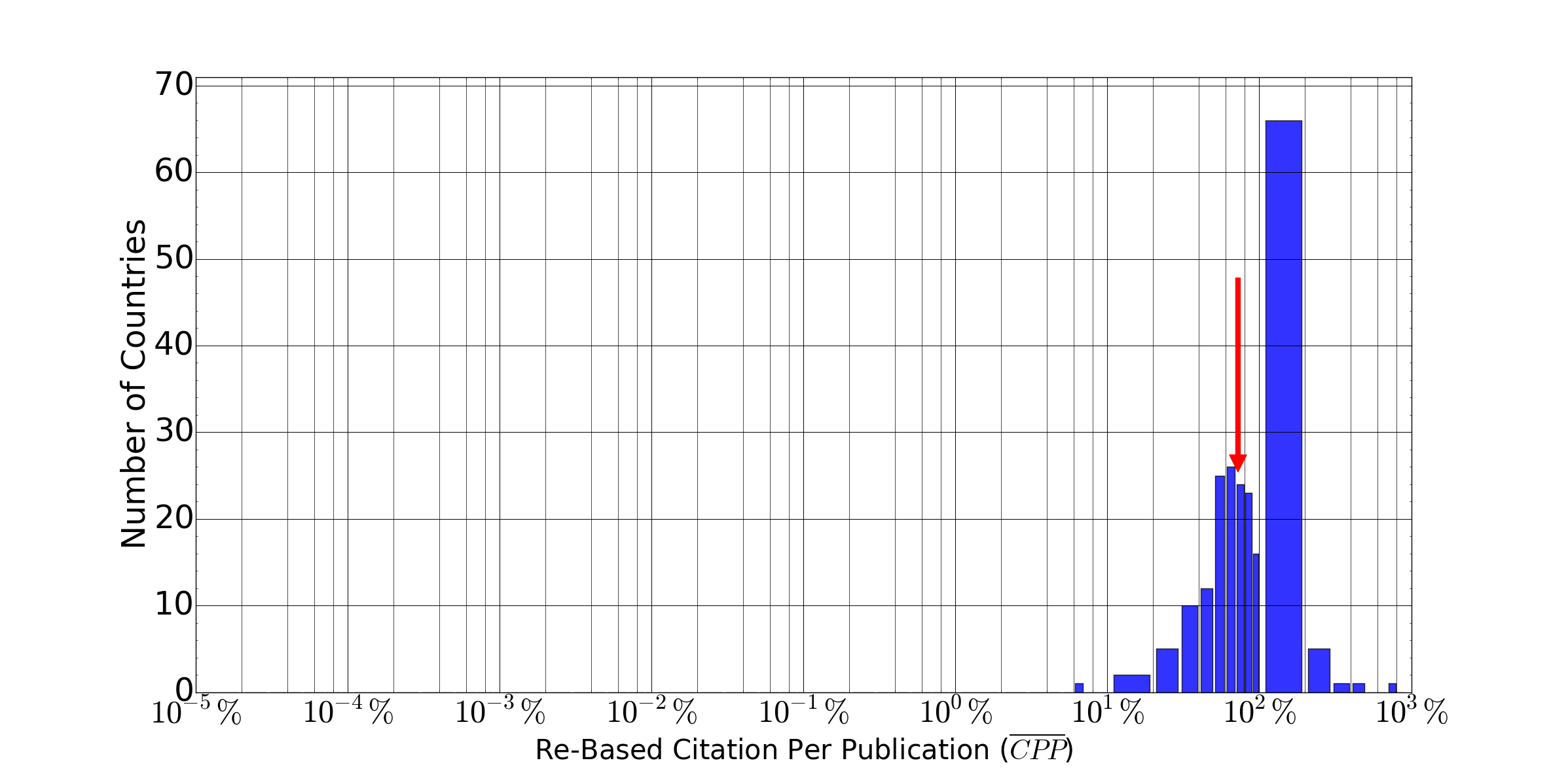}} 
  \caption{Ranking and distribution of countries in each of the indicators for the year 2014 (last year of the study)
  are shown on different rows, starting from the top
  quantity ($\overline{P}$), productivity by population ($\overline{PPP}$), productivity by GDP ($\overline{PPG}$), 
  impact ($\overline{C}$) and quality ($\overline{CPP}$).
  Rankings are presented on the left column in logarithmic scale. 
  South Africa is shown by red arrows in both ranking and distribution.
  }
  \label{RBCPP_ranking}
\end{figure}
Fig. \ref{RBCPP_ranking} shows ranking of South Africa 
among the countries of the world in all the five indicators
studied here. 
Furthermore, the distribution of the number of countries 
over the value of each indicator is displayed on the right column. 
Table \ref{ranking_SA} presents South Africa's rank in global context
which is given for each of the indicators in two different years, 
i.e. starting and ending years of the studied period.

One can infer from the rankings change and the shape of the distribution function that
the wider the distribution, the harder it is to change the ranking for countries. 
For instance, the width of the distribution of $\overline{C}$ and $\overline{CPP}$ are $10^6$ and $10^2$ respectively. 
The shift in South Africa's ranking in these two indicators are $-2$ and $-55$ respectively.
The drop in the rank of South Africa in the latter should not be interpreted as a huge failure
since the distribution of this indicator is so tight, and a change of rank is much easier to happen. 
In other words, the ranking should be cautiously interpreted by carefully examining the shape of the distribution of that indicator 
(more specifically on the width of the distribution). 
\begin{table}
 \begin{center}
\begin{tabular}{ |p{7cm}|p{2cm}||p{2cm}|p{2cm}||p{2cm}|  }
 \hline
 \multicolumn{5}{|c|}{South Africa's rank} \\
 \hline
 Name of the indicator& Concept & year 1996  &year 2014 &width of distribution function\\
 \hline
 Re-Based Publication ($\overline{P}$)  		&   Quantity      & 29 	 & 30  & $\simeq 10^6$ \\
 Re-Based Publication Per Population ($\overline{PPP}$) &   Productivity  & 53   & 82  & $\simeq 10^3$\\
 Re-Based Publication Per GDP ($\overline{GDP}$)        &   Productivity  & 54   & 35  & $\simeq 10^3$\\
 Re-Based Citation 	($\overline{C}$)  		&   Impact    	  & 28 	 & 30  & $\simeq 10^6$\\
 Re-Based Citation Per Publication ($\overline{CPP}$)   &   Quality    	  & 99 	 & 154 & $\simeq 10^2$\\
 \hline
\end{tabular}
\\[10pt]
\caption{ South Africa's rank is presented in two years, e.g. 1996 and 2014 along side the width of the distribution of each indicator in 2014.
One should carefully understand these numbers. As the distribution of the value for each of these indicators are different.
So, although the ranking are different for productivity and quality,
this should not be interpreted as a strong failure due to the narrow
width of the distribution of the indicators associated with these concepts,
in other words, smaller the width, easier to change rank. }
\label{ranking_SA} 
\end{center}
\end{table}

\section{Concluding notes} \label{Sec_Conclusion}
This study is motivated by the sharp increase in the demand for national level scientometrics, 
mostly originated from the global shift toward knowledge-based system of economy which emphasizes on 
science, research and innovation as a driving force.
Scientific performance and its four major aspects are focused upon. 
The ranking and the temporal evolution of South Africa in all the four aspects have been presented. 

In conclusion, South Africa has been successful enough to either steady increase or hold its position in the scientific world. 
However, there are other developing countries,
such as China, Turkey, Malaysia and Iran, 
which enjoyed much higher achievement than South African . 
South Africa's transition into a knowledge-based economy is far from being over and other developing countries
are going through the same transition, some of them more successful than South Africa.

\acknowledgments
This work is based upon research supported by the National Research Foundation 
and Department of Science and Technology.
Any opinion, findings and conclusions or recommendations expressed in this 
material are those of the authors and therefore the NRF and DST do not accept 
any liability in regard thereto.




\end{document}